\documentclass[12pt]{iopart}

\usepackage[utf8]{inputenc}
\usepackage{iopams}
\usepackage{amscd,amssymb}
\usepackage{bm}
\usepackage{epsfig}
\usepackage{graphicx}
\usepackage{xspace}
\usepackage{mathrsfs}
\usepackage{tabularx}

\usepackage{mathptmx}
\usepackage[T1]{fontenc}
\usepackage{url}
\expandafter\let\csname equation*\endcsname\relax
\expandafter\let\csname endequation*\endcsname\relax
\usepackage{amsmath,amssymb,amsfonts}

\usepackage{bm}
\usepackage{hyperref}
\usepackage{bbold}
\usepackage{xcolor}

\usepackage{tikz}

\newcommand \be  {\begin{equation}}
\newcommand \bea {\begin{eqnarray} \nonumber }
\newcommand \ee  {\end{equation}}
\newcommand \eea {\end{eqnarray}}

\renewcommand{\Re}{\mathop{\rm Re}}
\renewcommand{\Im}{\mathop{\rm Im}}

\newcommand{\algn}[1]{\begin{align} #1 \end{align}}
\newcommand{\sbeqs}[1]{\begin{subequations} #1 \end{subequations}}

\newcommand{\abs}{\ensuremath{\mathfrak{a}}}
\newcommand{\loc}{\ensuremath{\mathfrak{l}}}
\newcommand{\rat}{\ensuremath{\mathcal{R}}}
\newcommand{\ve}[1]{\boldsymbol{#1}}
\newcommand{\nn}{\nonumber}
\newcommand{\ete}{\ensuremath{\text{e}}}
\newcommand{\ed}{\ensuremath{\text{d}}}
\newcommand{\dd}[1]{\ensuremath{\tfrac{\text{d}}{\text{d} #1}}}

\newcommand{\eqnlab}[1]{\label{eq:#1}}
\newcommand{\seclab}[1]{\label{sec:#1}}
\newcommand{\figlab}[1]{\label{fig:#1}}
\newcommand{\eqnref}[1]{\eqref{eq:#1}}
\newcommand{\Eqnref}[1]{Eq.~\eqref{eq:#1}}
\newcommand{\Eqsref}[1]{Eqs.~\eqref{eq:#1}}
\newcommand{\secref}[1]{\ref{sec:#1}}
\newcommand{\Secref}[1]{Sec.~\ref{sec:#1}}
\newcommand{\figref}[1]{\ref{fig:#1}}
\newcommand{\Figref}[1]{Fig.~\ref{fig:#1}}
\newcommand{\Figsref}[1]{Figs.~\ref{fig:#1}}

\graphicspath{{./figs/}}

\begin{document}
\title{Density of reflection resonances in one-dimensional disordered Schr\"{o}dinger operators}

\author{Yan V. Fyodorov}
\address{King's College London, Department of Mathematics, London  WC2R 2LS, United Kingdom}
\ead{yan.fyodorov@kcl.ac.uk}

\author{Jan Meibohm}
\address{ Technische Universit\"{a}t Berlin, Institut f\"ur Physik und Astronomie, Fachgruppe Theoretische Physik, Hardenbergstr. 36, 10623 Berlin, Germany}
\ead{j.meibohm@tu-berlin.de}
\begin{abstract}
We develop an analytic approach to evaluating the density  $\rho ({\cal E},\Gamma)$ of complex resonance poles with real energies $\mathcal{E}$ and widths $\Gamma$ in the pure reflection problem from a one-dimensional disordered sample with white-noise random potential. We start with establishing a general link between the density of resonances and the distribution  of the reflection coefficient $r=|R(E,L)|^2$, where $R(E,L)$ is the reflection amplitude, at {\it complex} energies $E = {\cal E} +i\eta$, 
identifying the parameter $\eta>0$ with the uniform rate of absorption within the disordered medium.  We show that leveraging this link allows for a detailed analysis of the resonance density in the weak disorder limit. In particular, for a (semi)infinite sample, it yields an explicit formula for $\rho ({\cal E},\Gamma)$, describing the crossover from narrow to broad resonances in a unified way. Similarly, our approach yields a limiting formula for $\rho ({\cal E},\Gamma)$ in the opposite case of a short disordered sample, with size much smaller than the localization length. This regime seems to have not been systematically addressed in the literature before, with the corresponding analysis requiring an accurate and rather non-trivial implementation of WKB-like asymptotics in the scattering problem. Finally, we study the resonance statistics numerically for the one-dimensional Anderson tight-binding model and compare the results with our analytic expressions. 
\end{abstract}

\maketitle

\section{Introduction}\label{sec:intro}
\subsection{Anderson localization}
Disorder has profound impacts on the macroscopic properties of quantum systems. One of the most striking manifestations of this is Anderson localization~\cite{And58}, the suppression of wave propagation in a disordered medium due to quantum interference between multiple scattering paths. Originally introduced by P. W. Anderson in 1958 to describe the absence of diffusion of electrons in random lattices, Anderson localization has since emerged as a universal wave phenomenon observed in waves of light~\cite{Wie97}, matter~\cite{Bil08,Roa08}, and sound~\cite{Hu08}. In one and two dimensional quantum systems, even arbitrarily weak disorder leads to localization of all eigenstates. In higher dimensions, by contrast, a transition between localized and extended states, the Anderson (de)localization transition, may occur as disorder or energy is varied~\cite{Eve08}.

The standard way of modelling single-particle quantum propagation in a $d$-dimensional random medium $\mathcal{D}$ in general, and Anderson localization in particular, is based on the Hamiltonian
\be \label{Andcont}
H=-\frac{\hbar^2}{2m}\Delta_{\ve r}+V(\ve r)
\ee
with $\ve r \in {\cal D}\subset \mathbb{R}^d$ and appropriate (e.g. Dirichlet) conditions at the boundary of ${\cal D}$. This Hamiltonian combines the kinetic-energy Laplacian $-\Delta_{\ve r}$\footnote{Henceforth, we will choose {units} for which $\hbar^2/(2m)=1$ for brevity.} with a random potential $V(\ve r)$, $\ve r\in \mathbb{R}^d$. In the simplest case, $V(\ve r)$ is Gaussian, with zero mean $\langle V(\ve r) \rangle=0$, and correlation function $\langle V(\ve r) V(\ve r')\rangle=\frac{1}{2\pi \nu \tau}\delta(\ve r-\ve r')$, where $\nu$ denotes the mean density of energy levels per unit volume and $\tau$ stands for the mean-free time. Here and henceforth, the angular brackets $\left\langle\ldots\right\rangle$ denote averages over realizations of $V$.

The tight-binding analogue of Eq.~\eqref{Andcont} is given by the so-called Anderson Hamiltonian~\cite{And58}
\be\label{tightbinding}
H_\text{A}=\sum_{\ve r\in \mathfrak{L}}\,V_{\ve r}\left|\ve r\right\rangle\left\langle\ve r\right|+\sum_{\langle \ve r, \ve r'\rangle}t_{\ve r\ve r'}\left(\left.|\ve r\right\rangle\left\langle \ve r'\right|+\left|\ve r'\right\rangle\left\langle \ve r\right|\right),
\ee
where the second term includes a sum over nearest neighbours on the lattice $\mathfrak{L}$. The hopping parameters $t_{\ve r\ve r'}$ are assumed symmetric, $t_{\ve r\ve r'}=t_{\ve r'\ve r}$, to ensure that the Hamiltonian is self-adjoint. The form \eqref{tightbinding} can also be used for modelling a quantum particle on an arbitrary graph $\ve r\in {G}$. In this case, $t_{\ve r\ve r'}$ correspond to the weighed elements of the adjacency matrix of the graph ${G}$. The disordered tight-binding model is frequently called ``Anderson model''~\cite{And58} in the literature .

As mentioned above, the major single-particle wave-interference effect due to potential disorder is the Anderson localization phenomenon, which ensures that for strong enough disorder $V$, the eigenfunctions of $H$ or $H_\text{A}$ become localized, i.e., their magnitude decays exponentially on scales of the so-called localization length $\ell_\text{L}$. As a result, the classical single-particle diffusive dynamics, characterized in a random medium by a classical diffusion constant $D\sim E\tau$, is restricted to length scales of order $\ell_\text{L}$, which strongly impedes transport and transmission through the medium. In low-dimensional random media with $d\le 2$ the localization length remains finite for any degree of disorder. In higher dimensions, by contrast, the localization length $\ell_\text{L}$ increases with decreasing variance of $V(\ve r)${, for a fixed generic value of the energy $E$,} and eventually diverges.  At the corresponding critical disorder strength, the medium undergoes the aforementioned Anderson (de)localization transition, where, upon further decrease of the disordered potential variance {(at the chosen fixed $E$)} the eigenfunctions become extended. In the emerging delocalized phase, the eigenfunctions are randomly but uniformly spread over the whole sample, thus restoring diffusive dynamics inside the medium and enabling transport through the sample.  Various aspects of the Anderson localization problem, such as the associated anomalous quantum diffusion and multifractality of eigenfunctions in the vicinity of the delocalization transition, are a topic of active research~\cite{Eve08}.

\subsection{One-dimensional Anderson model}
The one-dimensional continuous version of the Anderson model plays {a prominent} role in the {theoretical} study of localization phenomena. Because of its relative simplicity and analytical tractability, it serves as the paradigmatic setting in which fundamental aspects of Anderson localization can be explored in full detail. The model is defined by the Hamiltonian
\be \label{eq:1dclosed}
H=-\frac{d^2}{d x^2}+V(x), \qquad x\in \mathcal{D}=[0,L]\,,
\ee
where the random potential $V(x)$ is a mean-zero white noise with correlation $\langle V(x) V(x') \rangle = D\delta(x - x')$, and appropriate conditions imposed at the boundaries $x=0$ and $x=L$ of $\mathcal{D}$.
In this setting, Anderson localization manifests itself in its most pronounced form: even for arbitrarily weak disorder, all eigenstates are exponentially localized in the system of infinite length $L\to \infty$, and many physical observables admit analytical treatment.

{An early systematic study of the one-dimensional model~\eqref{eq:1dclosed} was carried out by Halperin~\cite{Hal65}, who devised an efficient analytic approach to analyze the model~\eqnref{1dclosed}. Later developments} reached a major milestone with the Berezinskii approach~\cite{Ber74}, which enabled the calculation of key physical quantities in the semiclassical high-energy limit~\cite{Lif88}. These works demonstrated that the spatial correlations of wave functions decay exponentially, confirming that Anderson localization persists at all energies. Despite this, eigenfunctions at nearby energies exhibit significant fluctuations in both shape and spatial extent, motivating extensive studies of their statistical properties over the past three decades~\cite{Kol93,Tex00,Iva12}.

Beyond its physical relevance, the model~\eqref{eq:1dclosed} has become a cornerstone of rigorous mathematical analysis of {single-particle} disordered quantum systems, beginning with the foundational work of Gol'dshtein, Molchanov, and Pastur~\cite{Gol77}. The most complete mathematical understanding of its spectral statistics has been achieved only recently through a series of works by Dumaz and Labbé~\cite{Dum20,Dum23,Dum24a,Dum24b}.
\subsection{Wave scattering from one-dimensional disordered media}
Wave scattering from disordered media provides a complementary perspective on Anderson localization, connecting the spectral properties of disordered Hamiltonians to measurable transport phenomena. While the traditional “closed-system” formulation of Anderson localization, described above, focuses on the spectral statistics of the Hamiltonians $H$ and $H_\text{A}$, an equally important point of view arises from the study of open systems, where a disordered segment interacts with extended leads. In this scattering picture, localization manifests itself as the exponential suppression of transmission through the medium, which is a direct and experimentally accessible consequence of destructive interference between multiple scattering paths~\cite{She06}. In one dimension, such effects are particularly pronounced: even arbitrarily weak disorder exponentially impedes wave transmission, making the one-dimensional case a testing ground for fundamental aspects of localization and transport.

Technically, wave scattering in this context is described by embedding a disordered region of finite length  $L$ into an otherwise clean one-dimensional medium. The surrounding region, governed by the free Hamiltonian  $H_0=-\frac{d^2}{d x^2}$, supports incoming and outgoing plane waves of energy  $E=k^2>0$. If a unit-intensity wave is incident from the right, the stationary scattering state {for $x>L$} takes the form $u^{+}_k(x)=e^{ik(L-x)}+R(k,L)e^{ik(x-L)}${,} where 
 $R(k,L)$ is the reflection amplitude. In the region  $x<0$, the transmitted component reads $u^{-}_k(x)=T(k,L)e^{-ikx}$. The complex amplitudes $R(k,L)$ and $T(k,L)$ encode the reflection and transmission properties of the disordered segment and serve as central observables for characterizing localization in open systems.

The statistical behavior of these amplitudes has been the subject of extensive investigation, beginning with the seminal work of Gertsenshtein and Vasil’ev~\cite{Ger59}, who demonstrated, independently of Anderson, that $\lim_{L\to\infty}T(k,L)=0$, providing a scattering-based formulation of localization. The invariant imbedding method~\cite{Ram87} (see also Refs.~\cite{Ger59,Pap71,Lan73,Kum85}) offers a powerful framework for analyzing the stochastic evolution of $R$ and $T$ as functions of $L$. Alternative approaches, such as Berezinskii’s diagrammatic technique adapted to transmission and resistance statistics~\cite{Mel80,Abr81}, yield equivalent insights from a complementary perspective.

In electronic systems, the zero-temperature dimensionless DC conductance  $g$ of non-interacting spinless electrons is directly related to the transmission amplitude via  $g=|T(k,L)|^2$~\cite{Eco81,Fis81}. Experimentally, such scattering characteristics can be probed in classical wave analogues, for instance in microwave transmission experiments through random single-mode waveguides~\cite{Che17}, providing a direct testbed for the theoretical predictions of one-dimensional localization theory.

\begin{figure}
	\centering
	\includegraphics[]{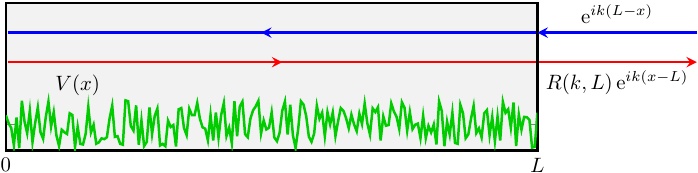}
	\caption{Sketch of the scattering setup. The incident wave (blue) enters the sample (grey) from the right, interacts with the disorder $V(x)$ (green), is back-reflected on its left end and eventually escapes the disordered region through its right end. Together with the reflecting left end of the sample, the disorder generates the scattered, outgoing wave (red), with reflection coefficient $R(k,L)$.}\label{fig:setup}
\end{figure}

The characteristics of the scattering problem is strongly dependent on the imposed boundary conditions. A natural choice requires continuity of the wavefunction $u(x)$ and its derivative $du/dx$ at the end points $x=0,L$ of the medium. These conditions ensure that the  Schr\"{o}dinger operator for the embedded piece of disordered medium is self-adjoint and admits both reflection from and transmission through the sample. An alternative choice is to require continuity at $x=L$ but impose Dirichlet boundary conditions $u(0)=0$ at $x=0$. This choice prohibits transmission and automatically ensures $T(k,L)=0$, giving rise to a pure scattering setup, as illustrated in \Figref{setup}. In this simple but non-trivial scattering scenario, an incident plane wave from the right scatters off the disordered medium and escapes the system again to the right.

The scenario of pure scattering from a disordered, finite, one-dimensional medium is the topic in this paper. Compared to the traditional closed-system formulation of Anderson localization, the  scattering setup changes the spectral properties of the associated operator  drastically: instead of discrete eigenvalues (energy levels) and associated eigenvectors (wave functions) one now deals with a continuous spectrum and plane wave scattering. Yet, some particular features in the scattering characteristics of {\it open} media can be treated as analogues of discrete eigenvalues of their {\it closed} counterparts. These features, known as {\it resonances}, are defined in the mathematical literature as poles of the meromorphic continuation of the resolvent of the associated operator from the lower half of the complex energy plane $\Im E<0$ to the whole complex plane~\cite{Dya19}. For a one-dimenisonal Schr\"odinger equation with generic non-random potential they have received considerable attention, see e.g. \cite{Zwo87,Fro97,Sim00,Kor04}.

In the physics literature, the same quantities are often defined as the poles of the scattering matrix in the lower half of the complex energy plane. In the present case of pure reflection, this scattering matrix is simply given by the reflection amplitude $R(E,L)$ as a function of the energy $E = k^2$ of the incident wave. This quantity is unimodular for real energies $E$ but acquires a nontrivial analytic structure for complex energies due to the occurrence of resonance poles. An important task is to characterize the statistics of the positions $E_n$ of these poles in the complex plane. Of particular interest are the imaginary parts $\Gamma_n = -\Im E_n$, commonly called {\it resonance widths}, whose values in the case of well-separated poles are associated with the inverse escape time of a quantum particle from the disordered region~\cite{Per98}. In the case of a high density of resonance poles (known in the literature as the regime of overlapping resonances), the association of resonance widths with escape times becomes more complicated due to interference effects mediated by non-orthogonality of resonance eigenfunctions. However, even in this case the knowledge of resonance widths statistics  may provide  useful insights into transport in disordered media and helps to reveal the temporal behavior of scattered signals, see e.g. \cite{Sav97}.

Another useful way to address temporal characteristics of the scattering process is via the Wigner time delay, defined as the energy derivative of the scattering phase shift. This quantity measures the typical time that incoming particles spend inside the disordered sample~\cite{Tex16}. Statistical properties of time delays in one-dimensional disordered media have been addressed in a similar framework in Refs.~\cite{Tex99,Oss00}. In fact, Wigner time delays and resonance poles are intimately, though nontrivially, related to each other -- a relation that has recently attracted renewed theoretical and experimental interest~\cite{Che21}.  
\subsection{Mean density of resonances}
A central characterization of the resonance-pole statistics is provided by the mean density $\rho({\cal E}, \Gamma)$ of imaginary parts $\Gamma_n$ of the resonance poles, whose real parts ${\cal E}_n = \Re E_n$ lie in the vicinity of an energy value ${\cal E}$, defined as  
\begin{equation}\label{resdendef}
 \rho ({\cal E},\Gamma):=
 \biggl\langle
 \sum_{n=1}^N \delta({\cal E}-{\cal E}_n)\,\delta\left(\Gamma-\Gamma_n\right)
 \biggr\rangle.
\end{equation}
Over the past two decades, $\rho ({\cal E},\Gamma)$ has been extensively studied in one-dimensional systems longer than the localization length, where the effects of Anderson localization are fully developed. A combination of analytical and numerical approaches, applied to both finite samples and the limit of infinite system size~\cite{Tit00,Ter00,Kun08,Fei09,Fei11,Gur12,Klo16,Her19}, has led to the consensus that the probability density decays as $\rho ({\cal E},\Gamma)\sim 1/\Gamma$ in the regime of narrow resonances, ${\Gamma_\text{L}} e^{-L/\ell_\text{L}} \ll \Gamma \ll {\Gamma_\text{L}}$, where $\ell_\text{L}$ denotes the localization length of eigenfunctions at energy ${\cal E}$, and ${\Gamma_\text{L}}$ is a typical resonance width at the same energy. This behavior can be traced back to the statistics of the inverse traversal time to the sample boundary and to the exponential localization of eigenfunctions for a typical disorder realization. The regime of ultranarrow resonances with $\Gamma \ll e^{-L/\ell_\text{L}}$ is governed by rare events and has been argued to display a log-normal form~\cite{Gur12}. Despite these advances, general nonperturbative results for the resonance-width density in the paradigmatic model~\eqref{eq:1dclosed}, covering the full range from narrow to broad resonances ($\Gamma \gg {\Gamma_\text{L}}$), remain largely unavailable.  

This situation contrasts with resonance scattering in quasi-one-dimensional wires, i.e., systems composed of many transversely coupled one-dimensional chains, where a detailed nonperturbative description of resonance densities for reflection from infinitely long wires has recently been achieved within the nonlinear sigma-model framework~\cite{Fyo24}.  

Here we present an analytic approach to evaluating $\rho ({\cal E},\Gamma)$ in the one-dimensional pure-reflection problem illustrated in Fig.~\ref{fig:setup}. Building upon and extending ideas from Refs.~\cite{Che21,Fyo24}, we establish a relation, Eq.~\eqref{main_rel}, between the resonance density $\rho ({\cal E},\Gamma)$ and the distribution $P(r,L)|_{{\cal E},\eta}$ of the reflection coefficient $r = |R(E,L)|^2$ at {\it complex} energies $E = {\cal E} + i\eta$. In this formulation, the parameter $\eta > 0$ represents a uniform absorption rate within the disordered medium. The distribution $P(r,L)|_{{\cal E},\eta}$ has been studied previously~\cite{Pra94,Fre97,Mis97} and satisfies a Fokker--Planck equation that can be solved by established methods~\cite{Ris89}. We show that exploiting this link enables an analytical treatment of the problem in various parametric regimes.  In particular, in the semiclassical (weak-disorder) limit $D \ll {\cal E}^{3/2}$, where ${\cal E} = k^2$, and for samples much longer than the localization length $\ell_\text{L} = 4{\cal E}/D$, we derive the following explicit formula for the resonance density:
\begin{equation}\label{rho_intro_long}
     \rho ({\cal E},\Gamma)=\frac{2k}{\pi D \Gamma}\left[1-\frac{8k\Gamma}{D}e^{\frac{8k\Gamma}D}\int_{\frac{8k\Gamma}{D}}^{\infty}e^{-t}\frac{dt}{t}\right],
\end{equation}
which goes considerably beyond existing results and describes, in a unified manner, the crossover from narrow ($\frac{D}{k}e^{-L/\ell_\text{L}}\ll \Gamma\ll \frac{D}{k}$) to broad ($\Gamma\gg \frac{D}{k}$) resonances; see the discussion following Eq.~\eqref{den_statexplicit}.
Moreover, our approach yields analytical results for the statistics of resonance widths in the opposite limit of very short samples ($L \ll \ell_\text{L}$), see Eq.~\eqnref{den_short}. {In this regime, localization is much less significant, and typical resonance widths are of the order of $\Gamma_\text{S}$, which scales inversely with the sample length $\Gamma_\text{S}=k/L$.} This regime appears not to have been systematically addressed in the literature and requires an accurate and rather nontrivial implementation of WKB-like asymptotics in the scattering problem.  

Finally, we study the resonance statistics numerically for discrete Anderson Hamiltonian~\eqref{tightbinding}, using exact and approximate methods similar to those in Refs.~\cite{Ter00,Her19}, and compare the numerical results with the theoretical predictions. Furthermore, we introduce an improved approximation scheme for scattering resonances, which approximates scattering resonances in the bulk of the spectrum as regular eigenvalues of an effective Hamiltonian. This approximation shares the advantages of a common approximation based on so-called ``parametric resonances''~\cite{Vin12,Her19}, but is significantly more accurate. We also show how the link we have established here provides new ways to compute the resonance density numerically.

\section{From the reflection coefficient to the density of complex poles}
To elucidate the relation between the reflection coefficient at complex energies and the density of resonance poles it is convenient to use the scattering setup for the 
discrete analogue of Eq.~\eqref{eq:1dclosed}, the Anderson model. To this end, we consider a semi-infinite lattice with sites labeled by $n=1,2,\ldots,\infty$ and with lattice spacing $a$.
The set of all sites will be subdivided into two subsets: The first, ``inner'' subset consists of $N$ sites with indices $1\le n\le N$, describing the disordered sample. This subset is described by the Anderson Hamiltonian
\algn{\label{eq:H_in}
	 H_\text{in} = \sum_{n=1}^{N}V_n|n\rangle\langle n|-\frac{1}{a^2} \sum_{n=1}^{N}\bigg(|n\rangle\langle n+1| -2 |n\rangle\langle n| + |n+1\rangle\langle n|\bigg)\,,
 \,}
where $V_1,\ldots,V_{N}$ denote Gaussian random variables with zero mean and correlation function
\algn{
	\langle V_n V_m\rangle = \frac{D}a\delta_{nm}\,.
}
The corresponding Schr\"odinger equation $H_\text{in}|\psi\rangle = E |\psi\rangle$ on the interval $[0,L]$ of length $L=Na$, replaces the continuum description associated with Eq.~\eqref{eq:1dclosed} by a system of $N$ coupled linear equations for the state vector $|\psi\rangle$.

By imposing Dirichlet boundary conditions $\langle0|\psi\rangle=0$ at the left end of the sample, the Schr\"odinger equation is written as
\be\label{innerSch}
	-\frac{1}{a^2}\bigg(\langle n+1|\psi\rangle - 2\langle n|\psi\rangle  + \langle n-1|\psi\rangle\bigg) + V_n\langle n|\psi\rangle = E \langle n|\psi\rangle\,,\quad  1\leq n\le N-1\,,
\ee
for all sites in the inner subset, except for the last one with $n=N$. For the lattice site $n=N$, located at the boundary of the ``inner'' region, we instead use a modified version of Eq.~\eqref{innerSch}, which reads
\be \label{boundary1}
	-\frac{1}{a^2}\bigg(t\langle N+1|\psi\rangle - 2\langle N|\psi\rangle  + \langle N-1|\psi\rangle\bigg) + V_{N}\langle N|\psi\rangle= E\langle N|\psi\rangle\,,
\ee
where $0\le t\le 1$ is a parameter that controls the coupling between the inner and outer parts of the system. In particular, for $t=0$ the system \eqref{innerSch}--\eqref{boundary1} of $N$ equations is closed and the corresponding spectral problem defines the set of $N$ real eigenvalues $E_\text{in}$, characterizing the spectrum of the Anderson Hamiltonian in Eq.~\eqref{eq:H_in}.

The second subset includes the ``outer'' sites labelled by indices $n\ge N+1$, and is used to provide the lattice analogue of the free Schr\"{o}dinger equation. The latter is given for $n\ge N+2$ by
\be \eqnlab{outerSch}
-\frac{1}{a^2}\bigg(\langle n+1|\psi\rangle - 2\langle n|\psi\rangle  + \langle n-1|\psi\rangle\bigg) = E \langle n|\psi\rangle\,,
\ee
At the boundary site $n=N+1$, located at the interface between the ``inner'' and ``outer'' regions, we instead have
\be \label{boundary2}
-\frac{1}{a^2}\bigg(\langle N+2|\psi\rangle - 2\langle N+1|\psi\rangle  + t\langle N|\psi\rangle\bigg) = E \langle N+1|\psi\rangle\,,
\ee
with the same control parameter $t$ as in Eq.~\eqref{boundary1}. By solving the free equation \eqnref{outerSch} in the form of a wave incident from the right and reflected back
 \be \label{reflectwave}
 	\langle n|\psi\rangle = \ete^{-ik(n-N-1)a}+R(k,N) \ete^{ik(n-N-1)a}\,,
 \ee
one obtains the dispersion relation
\be\label{disper}
	E = \frac{2}{a^2}\left[1- \cos(k a)\right]\equiv E(k)\,,
\ee
where we restrict to $k\in [0, \frac{\pi}{a}]$. In the continuum limit $a\to 0$, we recover $E(k)\to k^2$ thus matching the continuum case. From Eq.~\eqref{eq:H_in} we notice that for $t>0$ the system \eqref{innerSch}--\eqref{boundary1} of $N$ equations for the inner part of the lattice can be written as
\algn{\label{Kbound}
	\left[H_\text{in} -E(k){\bf 1}_N\right]|\psi_\text{in}\rangle=\frac{t \langle N+1|\psi\rangle }{a^2}|N\rangle\,,
}
where $|\psi_\text{in}\rangle =\sum_{n=1}^{N}\langle n|\psi\rangle |n\rangle$ and ${\bf 1}_N = \sum_{n=1}^{N}|n\rangle\langle n|$ denote $|\psi\rangle$ and the identity, respectively, restricted to the inner region.
Via projection onto $|N\rangle$, we find
\be\label{Kmatr}
	\frac{\langle N|\psi_\text{in}\rangle}{\langle N+1 |\psi\rangle}=-t \mathcal{K}, \quad \mathcal{K}=\frac{t}{a^2}\langle N|\left[E {\bf 1}_N-H_\text{in}\right]^{-1}|N\rangle\,.
\ee
On the other hand, the solution in Eq.~\eqref{reflectwave} implies
\be \label{re1}
	\langle N+1|\psi\rangle=1+R(k,N), \quad \langle N+2 |\psi\rangle =e^{-ika}+R(k,N)\,e^{ika}\,,
\ee
which, when substituted into Eq.~\eqref{boundary2} and using the dispersion relation \eqref{disper}, gives
\be \label{re2}
	\langle N|\psi_\text{in}\rangle=\frac{1}{t}\left(e^{ika}+R(k,N)\,e^{-ika}\right).
\ee
Finally, substituting Eqs.~\eqref{re1}--\eqref{re2} into Eq.~\eqref{Kmatr} yields a closed-form equation for the reflection amplitude at a given energy $E$ in the form
\be\label{reflect1}
	R(E,N)=-e^{2ika} \frac{1+t^2e^{-ika}\,\mathcal{K}}{1+t^2e^{ika}\,\mathcal{K}}\,, \quad E = \frac{2}{a^2}\left[1- \cos(k a)\right].
\ee
The right-hand side in Eq.~\eqref{reflect1} is manifestly unimodular for real values of $k$, as expected from flux conservation, however its pole structure is
somewhat hidden. To reveal the latter we use the identity $\det(\ve{1}+AB)=\det(\ve{1}+BA)$, which in the particular case of the rank-one matrix
$|N\rangle\langle N|$ implies
\be \label{detidet}
	 1+\alpha \mathcal{K}=\frac{\det\left(E {\bf 1}_N-H_\text{in}+\alpha|N\rangle \langle N|\right)}{\det\left(E {\bf 1}_N-H_\text{in}\right)}
\ee
valid for any complex $\alpha$. Equation~\eqref{detidet} allows to represent the reflection amplitude $R(E,N)$ as a ratio of determinants,
 \be\label{reflect2}
	R(E,N)=-e^{2ika} \frac{\det(E {\bf 1}_N-H_\text{eff}^\dagger)}{\det\left(E {\bf 1}_N-H_\text{eff}\right)}\,,
\ee
with the effective, non-Hermitian Hamiltonian
\algn{\eqnlab{Heffex}
	H_\text{eff}(k):=H_\text{in}-\frac{t^2e^{ika}}{a^2}\,|N\rangle \langle N|\,.
}
Although the Hamiltonian $H_\text{eff}(k)$ is restricted to the inner space, for $t>0$ it effectively describes the combined system of inner and outer sites. Furthermore, Eq.~\eqref{reflect2} implies that the complex poles for the reflection amplitude $R(E,N)$ coincide with the zeroes of the denominator on the right-hand side of Eq.~\eqref{reflect2}. These zeroes are given by the $N$ roots of the equation
\be\label{respoles_eq}
	\det\left(E {\bf 1}_N-H_\text{eff}\right)=0, \quad E = \frac{2}{a^2}\left[1- \cos(k a)\right].
\ee
For $t\to 0$ these roots  are all real and coincide with the eigenvalues $E_\text{in}$ of the inner self-adjoint Hamiltonian matrix $H_\text{in}$ in Eq.~\eqref{eq:H_in}. This observation is consistent with the uncoupling of Eqs.~\eqref{boundary1} and \eqref{boundary2}  in this limit, where the inner and outer parts of the system can be treated separately.

Evidently, the parameter $t>0$ controls the degree of coupling between the inner and outer parts. The coupling to the outer system perturbs the matrix $H_\text{in}$ inside the determinant in Eq.~\eqref{respoles_eq} by the complex rank-one perturbation $-t^2 a^{-2} e^{i k a}\,|N\rangle \langle N|$, which shifts all $N$ roots of the equation into the lower half of the complex energy plane, forming the {\it resonance poles}. Since the coefficient in front of the perturbation is itself energy-dependent, the exact roots are no longer solutions of a standard linear eigenvalue problem for any finite $N$, which requires non-standard tools in the numerical evaluation of exact resonances and calls for specialized approximation schemes. We discuss these methods in \Secref{numerics}.

Before we proceed, it is useful to note that, instead of defining resonances as $S$-matrix poles in the complex energy plane $E={\cal E}+i\eta$, one may equivalently define them as poles in the complex wavenumber plane $\kappa=\Re k+i\Im k$. As a function of $\kappa$, the reflection amplitude $R(\kappa,N)$, obtained by analytic continuation $k\to \kappa$ of Eq.~\eqref{reflect2}, takes the form $R(\kappa,N) = -D_N(-\kappa)/D_N(\kappa)$, where\footnote{Such representations are standard in the mathematical theory of scattering and are often used to analyze resonances for non-random potentials $V(x)$, see e.g.~\cite{Kor04}.} $D_N(\kappa) := e^{-i a \kappa}\det\!\left[E(\kappa){\bf 1}_N - H_\text{eff}(\kappa)\right]$. From this representation, the empirical density of zeros can be obtained using the Poincaré--Lelong identity by computing $\frac{1}{2\pi}\frac{\partial^2}{\partial \kappa \, \partial \overline{\kappa}} \ln|D_N(\kappa)|$. The resonances correspond to the zeros of the analytic function $D_N(\kappa)$ located in the quadrant $\Re \kappa > 0$, $\Im \kappa < 0$.

{In essence, s}uch an approach was used in Ref.~\cite{Fei09}, where the author {derived an expression for the} quantity $\left\langle \ln|D_N(\kappa)| \right\rangle$, averaged over white-noise potentials. {Similar formulas also appeared more recently in Ref.~\cite{Gas22}. From that starting point, the author of Ref.~\cite{Fei09} attempted to develop a theory for the density of resonance widths.} However, the resulting expression{s} turned out to be rather cumbersome, making it difficult to extract the full resonance-width {density} beyond the known $\Gamma^{-1}$ tail. This suggests that $|D_N(\kappa)|$ may not be the optimal quantity for analyzing the resonance density. Noting that $D_N(-\kappa)$ has no zeros for $\Im \kappa < 0$, and that the factor $e^{-2 i \kappa a}$ introduces none, one concludes that the resonance density can equivalently be obtained by differentiating the log-modulus of the reflection amplitude, $
\frac{1}{2\pi}\frac{\partial^2}{\partial \kappa \, \partial \overline{\kappa}} \ln|R(\kappa,N)|$.
 
We, however, find it technically easier and more transparent to work directly with resonances in the complex energy plane. To this end, we consider the reflection amplitude $R$ in \eqref{reflect2} as a function of the complex variable $E$. The resonances are then zeros of the denominator [see Eq,~\eqref{respoles_eq}] in the lower $E$ half-plane. Formally, neither the numerator nor the denominator are analytic functions of $E$ due to branch cuts starting at the spectral edges $\Re E = 0$ and $\Re E = 4a^{-2}$ and extending into the lower half-plane. 

In the continuum limit $N \gg 1$, however, the real parts of the resonance poles become dense, with the typical spacing between neighbouring real parts ${\cal E}_n$ remaining of order $(a^2 N)^{-1}$, a property unchanged by rank-one perturbations. Consider now the poles $E_n$ with real parts ${\cal E}_n$ in a {\it mesoscopic} energy window ${\cal E}_n \in ({\cal E}_0 - W/2, {\cal E}_0 + W/2)$ of width $(a^2 N)^{-1} \ll W \ll {\cal E}_0$, around a reference energy ${\cal E}_0$. Such a window contains many resonances, but the mean density of the real parts remains approximately constant within it. This condition is best fulfilled when the density varies least with energy, i.e., near the centre of the spectral support at ${\cal E}_0 = 2/a^2$, far from both spectral edges. 

We further assume that the widths satisfy $\Gamma_n \ll W \ll {\cal E}_0$, which is natural for resonant phenomena and will be justified {\it a posteriori} for the energy range relevant here. Under these conditions, the branch cuts in the energy plane become immaterial, and $R(E,N)$ can be treated as effectively meromorphic in the region containing the resonances. Writing $E = {\cal E} + i \eta$ with $\eta > 0$ and applying the Poincaré--Lelong identity, the counting measure of resonances in the chosen window can be written as
\be \label{lapl_ident_reflect}
\frac{1}{2\pi}\left(\frac{\partial^2}{\partial {\cal E}^2} + \frac{\partial^2}{\partial \eta^2}\right) \ln{|R({\cal E} + i\eta,N)|}
\approx \sum_{n} \delta({\cal E} - \Re E_n)\,\delta(\eta + \Im E_n).
\ee
Note that the sum on the right-hand side of Eq.~\eqref{lapl_ident_reflect} is nonzero because $\Im E_n < 0$. Averaging both sides over realizations of the random potential $\ve V = (V_1,\ldots,V_N)^{\sf T}$, we find that the mean resonance density $\rho({\cal E}, \Gamma)$ in the complex energy plane can be expressed through the statistics of the reflection coefficient $r(E,N) = |R(E,N)|^2$ for energies off the real axis as
\be \label{main_relA}
\rho({\cal E}, \Gamma) = \frac{1}{4\pi}\left(\frac{\partial^2}{\partial {\cal E}^2} + \frac{\partial^2}{\partial \eta^2}\right) \left\langle \ln r({\cal E} + i\eta,N) \right\rangle \Big|_{\eta = \Gamma}.
\ee
For $N \gg 1$, the energy dependence of $\left\langle \ln |r({\cal E} + i\eta,N)| \right\rangle$ for real ${\cal E}$ well inside the spectral support is expected to be smooth, varying only slowly with the mean level density. Hence, $\left\langle \ln |r({\cal E} + i\eta,N)| \right\rangle$ remains essentially constant across any mesoscopic window of real energies. By contrast, its dependence on $\eta$ is much sharper, occurring on a parametrically smaller energy scale related to typical resonance widths. We will verify these assumptions for the continuous one-dimensional system with white-noise potential below. The underlying mechanism, however, appears more general {and} leads us to retain only the second derivative with respect to $\eta$ in Eq.~\eqref{main_relA}, implying
\be \label{main_rel}
\rho({\cal E}, \Gamma) = \frac{1}{4\pi} \frac{\partial^2}{\partial \eta^2} \left\langle \ln r({\cal E} + i\eta,N) \right\rangle \Big|_{\eta = \Gamma}.
\ee
Equations~\eqref{main_relA} and \eqref{main_rel} are the main relations of this paper and the central starting point for all further analysis. Although our reasoning leading to Eq.~\eqref{main_rel} is new, the formula itself coincides with Eq.~(18) of Ref.~\cite{Fyo24}\footnote{Note the incorrect overall minus sign on the right-hand side of Eq.~(18) in Ref.~\cite{Fyo24}.}, and ultimately traces back to the relation between resonance poles and a complex generalization of the Wigner time delay discovered in Ref.~\cite{Che21}. In the next section, we recall the invariant imbedding approach and apply it to evaluate the average on the right-hand side of Eq.~\eqref{main_rel}.
\section{Invariant imbedding approach for the continuous model with white-noise potential}
As we already mentioned, the distribution $P(r,L)|_{{\cal E},\eta}$ of the reflection coefficient in the one-dimensional system with white-noise random potential has already been studied in the literature, see Refs.~\cite{Pra94,Fre97,Mis97}. For completeness, we briefly sketch below a derivation of the corresponding Fokker-Planck equation~\eqref{FPr}.

Complex energies $E= {\cal E}+i\eta$ are equivalent to replacing the potential $V(x)$ at every point $x$ by $V(x)-i\eta$.
Making such shift in the one-dimensional  model Eq.~\eqref{eq:1dclosed}, denoting ${\cal E}=k^2$ with real $k$, and following the invariant imbedding approach of Ref.~\cite{Ram87}, one finds that the reflected amplitude $R_k(x)$  for the interval open on the right satisfies a stochastic Ricatti equation of the form
\algn{
	\ed R_k(x)	= \left[2 i k R_k  -\frac{\eta}{2k}(1+R_k)^2 -\frac{D}{4k^2}(1+R_k)^3\right]\ed x - \frac{i\sqrt{D}}{2k} (1+R_k)^2 \cdot\ed W(x)\,,\label{Reqn}
}
for $0\leq x\leq L$. The initial condition $R_k(0) = 1$ implements the reflecting boundary condition at the left side of the disordered medium. In Eq.~\eqref{Reqn}, $\cdot\ed W(x)$ denotes the standard Brownian increment in It\^o convention\footnote{The additional cubic term in Eq.~\eqref{Reqn} compared with Ref.~\cite{Ram87} arises due this choice of convention.}~\cite{Gar09} with correlation function
\algn{
	\langle \ed W(x) \ed W(x') \rangle = \delta_{xx'}\ed x\,.
}
Similar considerations lead to the equation for the transmission coefficient $T_k(x)$, which reads in It\^o convention
\algn{
	\ed T_k(x) 	=& \left[ik - \frac{\eta}{2k}(1+R_k) - \frac{D}{4k^2}(1+R_k)^2 \right]T_k\,\ed x - \frac{i\sqrt{D}}{2k}(1+R_k)T_k \cdot\ed W(x)\,.\label{Teqn}
}
For reflective boundary conditions with $R_k(0)=1$, we must have $T_k(0)=0$, which immediately implies $T_k(x)=0$ for all $0\leq x\leq L$. However, Eq.~\eqref{Teqn} can be used to determine the localization and {absorption} length scales, as explained below.

Next, we express $R_k(x)$ and $T_k(x)$ through their moduli and phases as $R_k(x) = \sqrt{r_k(x)} \ete^{i\phi_k(x)}$ and $T_k(x) = \sqrt{t_k(x)} \ete^{i\varphi_k(x)}$.
For  $\eta=0$ the reflection amplitude must remain a pure phase $R_k(L)=e^{i\phi_k(L)}$, due to flux conservation, so that the reflection coefficient equals unity $r_k(L)= |R_k(L)|^2=1$ for all $L$.
Any finite absorption $\eta$ introduces a {absorption} length scale
\algn{\eqnlab{ellD}
	{\ell_\text{A}} = \frac{k}{\eta}\,,
}
which quantifies the exponential decay length of $|R_k|^2$ as function of $L$.

We consider here the so-called ``semiclassical'' case of large $k$ or weak disorder, characterized by $k\ell_\text{MFP}\gg1$, where $\ell_\text{MFP}$ is the mean free-path in the medium given by \cite{Lif88}
\be \eqnlab{ellMFP}
	\ell_\text{MFP} = \frac{\cal E}{D} =\frac{k^2}{D}\,.
\ee
In particular, $k\ell_\text{MFP}\gg1$ is equivalent to ${\cal E}\gg D^{2/3}$.

In the semiclassical limit, the phase $\phi_k$ rotates with a angular velocity (per unit length) $\omega \propto k$ to which, in the presence of absorption $\eta$, one needs to add a contribution from the inverse {absorption} scale ${\ell_\text{A}}^{-1}$. One then observes that when $\omega\ell_\text{MFP}\gg 1$ and $\omega{\ell_\text{A}}\gg 1$, the phase $\phi_k$ rotates on length scales much shorter than the characteristic lengths governing the change{s} of $r_k$ and $t_k$. Consequently, $\phi_k$ can be treated as a fast variable and be removed by adiabatic elimination~\cite{Lif88,Ris89}. This way, one obtains effective stochastic differential equations for $r_k$ and $t_k$ alone~\cite{Ram87}. The resulting effective equation for $t_k$ can, upon transforming into an equation for the logarithm, be expressed as
\algn{
	\ed \log t_k = -\left(\frac{\eta}{k} + \frac{D}{4k^2}\right)\ed x - \sqrt{\frac{D}{2k^2}}\sqrt{r_k}\cdot\ed W(x)\,.
}
Averaging over $W(x)$ gives the exponential decay of $t_k$ as a function of $x$
\algn{
	\left\langle\frac{\ed \log t_k}{\ed x}\right\rangle = \frac{\eta}{k} + \frac{D}{4k^2} = \ell^{-1}_{{\text{A}}} + \ell^{-1}_\text{L}\,,
}
where we identified the localization length
\algn{\eqnlab{ellloc}
	\ell_\text{L} = \frac{4 k^2}{D} = 4 \ell_\text{MFP}\,,
}
which is the characteristic length scale of the exponential decay of squared wave functions in the disordered medium in the absence of {absorption}.

To summarize, in the semiclassical limit the problem depends on three independent characteristic length scales: the sample length $L$, the localization length $\ell_\text{L}$~\eqnref{ellloc}, and {absorption} length ${\ell_\text{A}}$~\eqnref{ellD}. In the one-dimensional case considered here, the mean-free $\ell_\text{MFP}$~\eqnref{ellMFP} is proportional to the localization length and thus not independent. All length scales and their definitions are summarized in the first part of Tab.~\ref{tab:parameters}.

We continue by describing the dynamics of the reflection coefficient $r_k$. To lighten the notation, we suppress the index $k$ in the following. The effective equation for $r:=r_k$ is given by
\algn{\eqnlab{rIto}
	\ed r = \left[-2{\ell_\text{A}}^{-1} r + \ell_\text{L}^{-1}(1-r)^2\right]\,\ed x - \sqrt{2\ell_\text{L}^{-1}}\sqrt{r}(1-r)\cdot\ed W(x)\,.
}
with initial condition $r(0) = 1$. This equation can be straightforwardly simulated on a computer using discretization methods such as the Euler-Maruyama scheme, to obtain an approximation for the probability density $P(r,L)|_{{\cal E},\eta}$ of the reflection coefficient $r$ at sample length $L$. We will come back to this method in \Secref{num_main_rel}.

Furthermore, the It\^o equation~\eqnref{rIto} immediately gives rise to the Fokker-Planck equation
\be \label{FPr}
	\partial_{x} P(r,x) \equiv L_\text{FP}P(r,x) = \partial_r\left\{ r\left[2{\ell_\text{A}}^{-1} + \ell_\text{L}^{-1}\partial_r (1-r)^2\right]P(r,x)\right\}\,,
\ee
for $r\in[0,1]$ and $x\in[0,L]$, where $L_\text{FP}$ denotes the Fokker-Planck operator. Reflecting boundary conditions at the end of the sample are implemented by the initial condition
\be \label{ini}
\lim_{x\to 0} P(r,x)=\delta^-(r-1)\,.
\ee
Here, $\delta^\pm$ denote the right- and left-sided delta functions, for which $\int_{-\infty}^\infty\delta^\pm(x-a)f(x)\text{d}x = f(a^{\pm})$, respectively. The probability density for $r$ of a disordered medium of length $L$ is finally obtained by evaluating $P(r,x)$ at $x=L$, i.e.,  $P(r,L) = P(r,L)|_{{\cal E},\eta}$.

 The Fokker-Planck formulation is useful for extracting analytical results for $P(r,L)|_{{\cal E},\eta}$ in the limits of short and large sample size and it serves as a starting point for approaches to compute $\langle \log r\rangle$ in the main relation~\eqref{main_rel}.

\begin{table}
    \renewcommand{\arraystretch}{1.1}
	\begin{tabularx}{\textwidth}{|p{2.5cm}|X|p{7cm}|p{3.2cm}|}
        \hline
        type			&	symbol		&	meaning	&	definition \\
        \hline
        length scale	&	$L$			&	sample length		& -		\\
        				& $\ell_\text{L}$		& 	localization length 	& $\ell_\text{L} = \frac{4k^2}D$\\
        				& ${\ell_\text{A}}$	& 	{absorption} length 	& $\ell_\text{A} = \frac{k}\eta$\\
        				& $\ell_\text{MFP}$	& 	mean-free path	 	& $\ell_\text{MFP} = \frac{k^2}D=\frac{\ell_\text{L}}4$\\
        \hline
	energy scale	& {$\Gamma_\text{L}$}		& {typical resonance	width for long samples}	&	$\Gamma_\text{L} = \frac{k}{\ell_\text{L}} = \frac{D}{4k}$\\
				& {$\Gamma_\text{S}$}		& {typical resonance width for short samples}	&	$\Gamma_\text{S} = \frac{k}{L}$\\
				& $\Gamma_\text{min}$	& resonance width cutoff	&	$\Gamma_\text{min} = \Gamma_\text{L}\ete^{-\loc}$\\
				& $\mathcal{E}$	& typical energy		&	$\mathcal{E} = \Re E=  k^2$\\
                & $\eta$	&  absorption rate	&	$\eta= \Im E$\\
	\hline
	dimensionless			&	${\mathcal{G}}$	&	{resonance width relative to $\Gamma_\text{L}$} 	&	$\mathcal{G} = \frac{\Gamma}{\Gamma_\text{L}} = \frac{4k\Gamma}{D}$	\\
						&	${\mathfrak{G}}$		&	{resonance width relative to $\Gamma_\text{S}$}	 	&	$\mathfrak{G}  = \frac{\Gamma}{\Gamma_\text{S}} =\frac{L\Gamma}{k}= \loc\mathcal{G}$	\\
						&	{$\mathfrak{a}$}		&	{degree of absorption}	 	&	$\mathfrak{a} = \frac{L}{\ell_\text{A}} = \frac{L\eta}{k}$	\\
						&	{$\mathfrak{l}$}		&	{degree of localization}		&	$\mathfrak{l} = \frac{L}{\ell_\text{L}}=\frac{LD}{4k^2}$\\
						&	{$\mathcal{R}$}	&	{absorption-localization ratio} 	&	$\mathcal{R} =\frac{\ell_{L}}{\ell_\text{A}}=\frac{4k\eta}{D}= \frac{\mathfrak{a}}{\mathfrak{l}}	$\\
	\hline			
        \end{tabularx}
        \caption{Summary of the main parameters and scales.}\label{tab:parameters}
\end{table}

\subsection{Mean of the logarithm}\seclab{meanlog}
According to the main relation~\eqref{main_rel}, the statistics of resonance widths is connected to the mean of the logarithm $\langle\log r\rangle$ of the reflection coefficient $r$. Starting from the Fokker-Planck equation~\eqref{FPr} there are multiple ways in which we can leverage this result. 

First, in the limit of (semi) infinite sample $L\to \infty$, which physically corresponds to lengths such that $x=L\gg\ell_\text{L},{\ell_\text{A}}$, the solution to the Fokker-Planck equation tends to the stationary solution $P_\text{st}(r)$ satisfying 
\algn{
	L_\text{FP}P_\text{st}(r)= \partial_r\left\{ r\left[2{\ell_\text{A}}^{-1} + \ell_\text{L}^{-1}\partial_r (1-r)^2\right]P_\text{st}(r)\right\} = 0\,,
}
which after simple manipulations gives
\algn{\label{stationary}
	P_\text{st}(r) = \frac{2{\rat}}{(1-r)^2}\text{e}^{-\frac{2{\rat} r}{1-r}}\,,
}
where ${\rat} = \ell_\text{L}/{\ell_\text{A}}  = (4k\eta)/D$ {denotes the absorption-localization ratio. In a long sample, $L\gg{\ell_\text{A}},\ell_\text{L}$, $\rat$ characterizes the relative importance of absorption and localization. The solution~\eqref{stationary}} is equivalent to the well-known stationary distribution of the reflection coefficient $r$ in a semi-infinite disordered system with absorption~\cite{Pra94}. Together with the main relation~\eqref{main_rel}, Eq.~\eqref{stationary} allows us to obtain the distribution of resonances widths in the infinite-sample limit. We consider this case in Sec.~\ref{lsl}.
\subsubsection{Direct computation of $\langle\log r\rangle$} Instead of first computing $P(r,x)$ from Eq.~\eqref{FPr} and then deriving $\langle\log r\rangle$, a more direct way is provided by using the Fokker-Planck equation~\eqref{FPr} to obtain an expression for $\langle\log r\rangle$. We multiply Eq.~\eqref{FPr} by $\log r$ and integrate over $r$ to find
\algn{\label{dlog}
		\dd{x} \langle \log r\rangle(x) &= \int_0^1\!\!\ed r\,\log r\,\partial_r r \left[2{\ell_\text{A}}^{-1} + \ell_\text{L}^{-1}\partial_r(1-r)^2\right]P(r,x)\,,\nn\\
		&= -\log r J(r,x)\big|_{0^+}^{1^-} - \int_0^1\!\!\ed r \left[2{\ell_\text{A}}^{-1} + \ell_\text{L}^{-1}\partial_r(1-r)^2\right]P(r,x)\,,
}
where $J(r,x) = -r \left[2{\ell_\text{A}}^{-1} + \ell_\text{L}^{-1}\partial_r(1-r)^2\right]P(r,x)$ denotes the probability flux at $r$ and $x$. The boundary terms in Eq.~\eqref{dlog} vanish, because of the no-flux condition $J(r,x)\big|_{0^+,1^-}=0$ at the boundaries. Using the normalization of $P(r,x)$ we find
\algn{\eqnlab{logrrel}
	\dd{x} \langle \log r\rangle(x)  = -2{\ell_\text{A}}^{-1} - \ell_\text{L}^{-1}(1-r)^2P(r,x)\big|_{0^+}^{1^-} =\ell_\text{L}^{-1} P(0^+,x)-2{\ell_\text{A}}^{-1}\,.
}
Integrating $x$ from $x=0$ to the sample length $x=L$, using $P(r,0)=\delta^{-}(r-1)$ we obtain
\algn{\eqnlab{logrint}
	\langle \log r\rangle = \ell_\text{L}^{-1}\int_0^{L}\!\!\ed x\,P(0^+,x) - 2L{\ell_\text{A}}^{-1}\,.
}
This expression is useful, because it connects $\langle \log r\rangle$ to $P(r,x)$ for a single $r=0^+$, but for all values of $x\in[0,L]$. In particular, in \Secref{ltwkb}, we determine $P(0^+,x)$ in the short-sample limit using methods of singular perturbation theory, which yields an explicit expression for $\langle \log r\rangle$.
\subsubsection{WKB method}
In the so-called short-sample (or ballistic) regime, characterized by the condition $L\ll\ell_\text{L}$~\cite{Her19} the particle wave function is only weakly affected by the disorder and spread over the entire medium. As a consequence, it strongly interacts with the boundaries and we expect the resonances to be short-lived, with large widths $\Gamma$ of the order of $\Gamma_\text{S}=k/L$. Starting from the Fokker-Planck equation~\eqref{FPr}, we develop a theory for the distribution of resonance widths in this limit, which, to the best of our knowledge, has not yet been discussed in the literature.

First, we find convenient to dedimensionalize the Fokker-Planck equation using the sample length $L$, so that $L\to\tilde L=1$ and $x\to\tilde x = x/L$, with the small parameter ${\loc} = L/\ell_\text{L} \ll 1$ {measuring the degree of localization in a finite sample}. In these coordinates the Fokker-Planck equation~\eqref{FPr} reads
\algn{\eqnlab{FPrwn}
	\partial_{\tilde x} \tilde P(r,\tilde x) = \partial_r r\left[2{\abs} + \loc\partial_r(1-r)^2 \right]\tilde P(r,\tilde x)\,,
}
where $\tilde P(r,\tilde x) \equiv P(r,x)$ and ${\abs} = L/{\ell_\text{A}} = L\eta/k$ denotes the {degree of absorption, i.e., the sample length $L$ relative to the absorption length $\ell_\text{A}$}.

Before we proceed, we briefly summarize the relevant dimensionless parameters in the limits of short and large sample length. In the long-sample limit, the stationary distribution $P_\text{st}(r)$ in Eq.~\eqref{stationary} depends solely on the dimensionless {absorption-localization ratio} ${\mathcal{R}}=\ell_\text{L}/{\ell_\text{A}}$. By contrast, for short samples, the problem depends on both the dimensionless {degree of localization} ${\loc} = L/\ell_\text{L}$, and on the {degree of absorption} ${\abs} =L/{\ell_\text{A}}$. All dimensionless variables are summarized in the lower part of Tab.~\ref{tab:parameters}.

To find an approximate solution of the Fokker-Planck equation~\eqnref{FPrwn} in the limit ${\loc}\ll1$, we employ the WKB Ansatz
\algn{\eqnlab{wkbans}
	\tilde P(r,\tilde x) \sim \frac{1}{\sqrt{2\pi{\loc}}}G(r,\tilde x)\ete^{-S(r,\tilde x)/{\loc}}\,,\qquad 0\leq \tilde x \leq1\,.
}
Substituting the latter to \Eqnref{FPrwn} yields, to leading order in ${\loc}$, the Hamilton-Jacobi equation
\algn{\eqnlab{hjeqn}
	0=\partial_{\tilde x} S(r,\tilde x) + \mathcal{H}[r,\partial_{\tilde x} S(r,\tilde x)]\,,
}
with the Hamiltonian function
\algn{\eqnlab{dynham}
	\mathcal{H}(q,p) = -2{\abs} qp + q(1-q)^2 p^2\,.
}
Equation~\eqnref{hjeqn} can be solved by characteristics $[q(s),p(s)]_{0\leq s\leq \tilde x}$ that obey the Hamilton equations
\sbeqs{\eqnlab{hameom}
\algn{
	\dd{s}q =& \partial_p {\mathcal H}(q,p) = 2 q \left[p (1-q)^2-{\abs} \right]\eqnlab{qeqn}\,,\\
	\dd{s}p =& -\partial_q {\mathcal H}(q,p) = -p \left[p (1-q) (1-3 q)-2 {\abs}\right]\,,
}
}
with boundary conditions $q(\tilde x) = r$ and $q(0) = 1$, where the latter condition implements the reflective nature of the left-hand side of the sample. From the characteristics $[q(s),p(s)]_{0\leq s\leq \tilde x}$, $S(r,\tilde x)$ is obtained as
\algn{\eqnlab{actionint}
	S(r,\tilde x) = \int_0^{\tilde x}\!\!\!\ed {s} \left[p\dd{s} q - \mathcal{H}(q,p) \right]\,.
}
The action $S(r,\tilde x)$ gives the exponential contribution the probability density $\tilde P(r,\tilde x)$.

The prefactor $G(r,\tilde x)$ in \Eqnref{wkbans} can be computed as follows: We write $G(r,\tilde x) = \ete^{-T(r,\tilde x)}$ and derive an equation for $T(r,\tilde x)$ from the next-to-leading order in ${\loc}$, giving
\begin{multline}
	0=2 {\abs} +\partial_{\tilde x}T(r,\tilde x)+ \partial_rT(r,\tilde x) \left[2 (1-r)^2 r \partial_rS(r,\tilde x)-2 {\abs}  r\right]\\
	-r (1-r)^2 \partial_r^2S(r,\tilde x)+\left[(6-5 r) r-1\right] \partial_rS(r,\tilde x)\,.
\end{multline}
Evaluated along the characteristics $[q(s),p(s)]_{0\leq s\leq \tilde x}$, and using \Eqnref{hameom}, we rewrite this equation as
\algn{\label{Teqn_new}
	0=&\partial_{s}T[q(s),s]+ \partial_qT[q(s),s] \dd{s}q+2 {\abs} -q (1-q)^2 Z(\tilde x)+\left[(6-5 q) q-1\right] p\,,
}
where $Z(s) = \partial_q^2 S[q(s),s]$ denotes the curvature of $S$ along the characteristic. An equation for $Z(s)$ is obtained by taking a double $q$-derivative of the Hamilton-Jacobi equation~\eqnref{hjeqn} and evaluating it along the characteristic. This way, we find the following Ricatti equation for $Z(s)$
\algn{
	\dd{s}Z(s)		&= - \partial^2_q {\mathcal H}(q,p) - 2\partial_q\partial_p {\mathcal H}(q,p)Z - \partial^2_p {\mathcal H}(q,p)Z^2\,,\nn\\
				&= 2p^2\left(2-3q\right) -2\left[2p(1-q)(1-3q)-2{\abs}\right]Z(s) -2q\left(1-q\right)^2Z(s)^2\,. \eqnlab{Zeqn}
}
The initial condition $P(r,0) = \delta^{-}(r-1)$ forces $Z(s) \to \infty$ as $s\to 0$. Finally, we identify the first two terms in Eq.~\eqref{Teqn_new} as the total derivative of $T$ along $q(s)$ and find
\algn{
	\dd{s} T(s) =& -2{\abs} + q(1-q)^2 Z(s) + (1-q)(1-5 q) p\,,\nn\\
			 =& \frac{p^2(2-3q)}{Z} - \dd{s}\log\sqrt{\text{e}^{2{\abs} s}qZ}\eqnlab{Teqs}\,,
}
where $T(s) = T[q(s),s]$ and we used Eqs.~\eqnref{qeqn} and \eqnref{Zeqn} in the second step. Integrating \Eqnref{Teqs} from $s=0$ to $s=\tilde x$, and choosing $T(0)=-\log\sqrt{q(0)Z(0)}\to-\infty$, leads us to the expression
\algn{\eqnlab{Teqn2}
	T(\tilde x) = \int_0^{\tilde x}\ed {s} \frac{p^2(s)[2-3q(s)]}{ Z(s)} -\log\sqrt{\text{e}^{2{\abs} \tilde x}q(\tilde x) Z(\tilde x)}\,.
}
The initial condition for $T$ in \Eqnref{Teqn2} is chosen such that $\tilde P(r,\tilde x)$ in \Eqnref{wkbans} obtains the proper normalization. This can be seen by considering the dynamics of the minimum of $S(r,\tilde x)$: The characteristics $[q^*(s),p^*(s)]_{0\leq s\leq \tilde x}$ along this minimum satisfy $p^*(s) = \partial_q S[q^*(s),s] = 0$ and $q^*(s) = \text{e}^{-2{\abs} s}$, which gives $S[q^*(\tilde x),\tilde x] = 0$. The normalization condition reads
\algn{
	1&=\int_0^1\!\!\ed r \tilde P(r,\tilde x) \sim 	\int_0^1\!\!\ed r \,\frac{\text{e}^{-\frac{S(r,\tilde x)}{{\loc}}-T(r,\tilde x)}}{\sqrt{2\pi{\loc}}}\sim \frac{\text{e}^{-T[q^*(\tilde x),\tilde x]}}{\sqrt{2\pi{\loc}}}
	\int_0^1\!\!\ed r \,\text{e}^{-\partial_q^2 S[q^*(\tilde x),\tilde x](r-q^*(\tilde x))^2/(2{\loc})}\,,\nn\\ 
	&\sim \frac{\text{e}^{-T[q^*(\tilde x),\tilde x]}}{\sqrt{\partial_q^2S[q^*(\tilde x),\tilde x]}} = \text{e}^{-\left\{T[q^*(\tilde x),\tilde x]+\log\sqrt{\text{e}^{2{\abs} \tilde x}q^*(\tilde x)\partial_q^2S[q^*(\tilde x),\tilde x]}\right\}}\,.
	\eqnlab{normalization}
}
By substituting $[q^*(s),p^*(s)]_{0\leq s\leq \tilde x}$ into \Eqnref{Teqn2}, and using $Z^*(\tilde x) = \partial_q^2 S[q^*(\tilde x),\tilde x]$, we confirm that the initial condition chosen in \Eqnref{Teqn2} is consistent with the normalization condition~\eqnref{normalization}.
\subsubsection{Gaussian approximation}
In addition to the normalization condition~\eqnref{normalization}, the dynamics of the minimum $[q^*(s),p^*(s)]_{0\leq s\leq \tilde x}$  also gives rise to a simple Gaussian approximation that describes the probability distribution close to $q^*(\tilde x)$ for small ${\loc}$, so that $|r-q^*(s)|\lessapprox {\loc}$, whenever $q^*(\tilde x)$ is sufficiently far away from the left boundary, $q^*(\tilde x)\gg {\loc}$. To obtain this approximation, we expand $S(q^*+r,\tilde x)$ for small $r$
\algn{\eqnlab{SGauss}
	S[q^*(\tilde x)+r,\tilde x] \sim S[q^*(\tilde x),\tilde x] + \partial_q S[q^*(\tilde x),\tilde x] (r-q^*(\tilde x)) + \frac12 \partial^2_q S[q^*(\tilde x),\tilde x] (r-q^*(\tilde x))^2\,.
}
The first two terms on the right-hand side of \Eqnref{SGauss} vanish because $p^*(\tilde x)=0$. As for the third term, $Z^*(\tilde x) =S[q^*(\tilde x),\tilde x]$ obeys the dynamics [cf. \Eqnref{Zeqn}]
\algn{
	\dd{s}Z^*(s)= 4{\abs}\, Z^*(s) -2\text{e}^{-2{\abs} s}\big(1-\text{e}^{-2{\abs} s}\big)^2 Z^*(s)^2\,,
}
with solution
\algn{
	Z^*(s) =  \frac{{\abs}}{2}\frac{\ete^{4{\abs} s}}{\sinh(2{\abs} s)-2{\abs} s}\,.
}
With the expression $T^*(\tilde x) = -\log\sqrt{Z^*(\tilde x)}$ [\Eqnref{Teqn2}] for the prefactor, the Gaussian approximation of $\tilde P(r,1) = P(r,L)$ reads
\algn{\eqnlab{pdfGauss}
	P(r,L) \approx \mathcal{N}\sqrt{\frac{Z^*(1)}{2\pi{\loc}}} \exp\left[-\frac{Z^*(1)}{2{\loc}}(r-\ete^{-2{\abs}})^2\right] I_{[0,1)}(r)\,,
}
where $\mathcal{N}$ denotes a normalization factor and $I_{S}(r)$ denotes the indicator function on the interval $S$, which is equal to unity when $r\in S$, and zero otherwise. The factor $\mathcal{N}$ reads
\algn{
	\mathcal{N}^{-1} = 	\frac12\left\{\text{erf}\left[\sqrt{\frac{Z^*(1)}{2{\loc}}}\ete^{-2{\abs}}\right] - \text{erf}\left[\sqrt{\frac{Z^*(1)}{2{\loc}}}(\ete^{-2{\abs}}-1)\right]\right\}\,.
}
\begin{figure}
	\includegraphics[width=\linewidth]{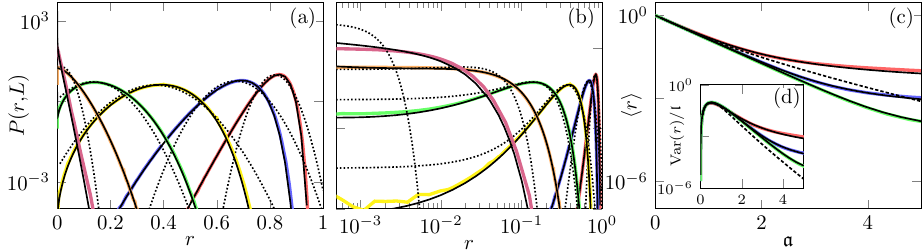}
	\caption{Comparison between numerical simulations of the stochastic dynamics~\eqnref{rIto} ($10^8$ realizations) and theory. (a) Probability distribution $P(r,L)$ for ${\loc}=10^{-1}$ as function of $r$ from numerical simulations for ${\abs} = 0.1$ (red), ${\abs} = 0.2$ (blue), ${\abs} = 0.5$ (yellow), ${\abs} = 1$ (green), ${\abs} = 2$ (orange), and ${\abs} = 5$ (purple) together with the WKB approximation (solid black lines) and the Gaussian approximation~\eqnref{pdfGauss} (dotted lines). (b) Same as in (a) but on a logarithmic $r$-axis. (c) $\langle r\rangle$ from numerical simulations as function of ${\abs}$ for ${\loc}=10^{-1}$ (red), ${\loc}=10^{-2}$ (blue), and ${\loc} = 10^{-3}$ (green), together with the WKB {approximation} (solid black lines) and from the Gaussian approximation (dotted lines). (d) Variance $\text{Var}(r)$, from the same data as in (c).}\figlab{distributions}
\end{figure}

To summarize,  \Eqnref{Zeqn} together with the Hamilton equations of motion~\eqnref{hameom} allows us to determine $P(r,L)\sim \tilde P(r,1)$ through the WKB Ansatz~\eqnref{wkbans}. The Hamilton equations~\eqnref{hameom} when combined with the appropriate boundary conditions $q(0)=1$ and $q(\tilde x) =r$  constitute a two-point boundary value problem, which we find challenging to solve analytically. Instead, we apply a shooting method~\cite{Pre86} to solve the system of \Eqsref{hameom} and \Eqnref{Zeqn} numerically. The resulting trajectories are then used to perform the numerical integrations in \Eqsref{actionint} and \Eqnref{Teqs}, to finally determine $P(r,L)$ using \Eqnref{wkbans}.

Figure~\figref{distributions}(a) shows $P(r,L)$ obtained from numerical simulations of \Eqnref{rIto} together with the WKB~\eqnref{wkbans} and the Gaussian approximations~\eqnref{pdfGauss}. We observe that as {the degree of absorption $\abs$} increases, the maximum of $P(r,L)$ moves towards the boundary at $r=0$. This is intuitive, because increasing ${\abs}$ corresponds to larger {absorption} within the sample, and thus a smaller reflection coefficient. When ${\abs}$ is large, enough $P(r,L)$ eventually merges with the boundary, generating a single maximum at $r=0$. Figure~\figref{distributions}(b) shows this merging process on a logarithmic $r$ scale. In all observed cases, we find that the WKB approximation~\eqnref{wkbans} provides a rather accurate representation of $P(r,L)$ for values ${\loc}\lessapprox 1$, i.e., for sample lengths up to the order of the localization length $\ell_\text{L}$. By contrast, the Gaussian approximation~\eqnref{pdfGauss} is quite accurate only for small values of ${\abs}$ and becomes increasingly less accurate for larger ${\abs}$ when the maximum of $P(r,L)$ approaches the boundary at $r=0$. This behavior is echoed in the mean $\langle r\rangle$ and the variance $\text{Var}(r) = \langle r^2\rangle - \langle r\rangle^2$ shown in \Figsref{distributions}(c) and \figref{distributions}(d), respectively. Here again, the WKB method provides an {accurate} approximation, while the Gaussian approximation fails for ${\abs}\gtrapprox1$.
\subsection{Laplace-transformed WKB}\seclab{ltwkb}
The WKB approach developed thus far allows us to compute the probability density $P(r,L)$ numerically. In this Section, we show how to use a Laplace-transformed formulation of the WKB theory, that allows us to obtain analytical results in the short-sample ballistic limit. To this end, we start with introducing the (rescaled) Laplace transform of $\tilde P(r,\tilde x)$:
\algn{\eqnlab{laptransf}
	\hat P(r,s) = \int_{0}^\infty\ed \tilde x \, \ete^{-\frac{s\tilde x}{{\loc}}}\tilde P(r,\tilde x )\,,
}
where we assume that $\tilde P(r,\tilde x)=0$ for $\tilde x>1$, i.e., outside the medium. The Laplace-transformed probability $\hat P(r,s)$ is found to satisfy
\algn{\eqnlab{lapfp}
	-\delta^-(1-r) + \frac{s}{{\loc}} \hat P(r,s) = \partial_r r \left[2{\abs} + {\loc}\partial_r(1-r)^2\right]\hat P(r,s)\,.
}
To determine the boundary condition for $\hat P(r,s)$ we integrate \Eqnref{lapfp} from $1-\zeta$ to $1$, where $\zeta\to0^+$. This gives
\algn{\eqnlab{boundcond}
	\hat P(1^-,s) = \frac1{2{\abs}}\,,
}
where we have used $\hat{P}(1,s) = 0$. After inverse Laplace transform, \Eqnref{boundcond} corresponds to $\tilde P(1^-,\tilde x) = (2{\abs})^{-1}\delta^+(\tilde x)$. Although unintuitive at first sight, this expression is consistent with the observation that at $r=1^-$ the dynamics~\eqnref{rIto} is purely deterministic, so that $\tilde P(1^-,\tilde x) = \delta^-[q^*(\tilde x)-1] = \delta^-(\ete^{-2{\abs}\tilde x}-1) =  (2{\abs})^{-1}\delta^+(\tilde x)$.

Furthermore, the reflective boundary condition at $r=0$ leads to a vanishing probability flux $\tilde J(r,\tilde x)$ at $r\to0^+$. For the Laplace transform~\eqnref{laptransf} the vanishing flux condition implies
\algn{\eqnlab{noflux}
	\lim_{r\to0^+} r\left[\partial_r\hat P(r,s)-\frac{2{\abs}}{{\loc}} \hat P(r,s)\right] = 0\,,
}
which imposes an additional constraint on $\hat P(r,s)$.

In order to solve the Laplace-transformed equation~\eqnref{lapfp} approximately, we employ the WKB Ansatz
\algn{\eqnlab{wkbssol}
	\hat P(r,s) \sim\frac{1}{\sqrt{2\pi{\loc}}}\hat G(r,s)\text{e}^{-\hat S(r,s)/{\loc}}\,,\qquad \hat G(r,s) = \ete^{-\hat T(r,s)}\,,
}
which gives the following equations for $S(r,s)$ and $T(r,s)$:
\sbeqs{\eqnlab{steqs}
\algn{
	s&= \partial_r \hat S(r,s)r\left[-2 {\abs}  +(1-r)^2 \partial_r \hat S(r,s)\right] = \mathcal{H}[r,\partial_r \hat S(r,s)]\,, \eqnlab{S0seqns}\\
	2 {\abs}  r \partial_r \hat T(r,s)&=2 {\abs} -\partial_r \hat S(r,s) \left[(1-r)(1-5r)-2 r (1-r)^2 \partial_r \hat T(r,s)\right]-r (1-r)^2 \partial^2_r \hat S(r,s)\,.
}
}
This pair of non-linear equations has two solutions $[\hat S^\pm(r,s), \hat T^\pm(r,s)]$. An expression for $\hat P(r,s)$ is then obtained by the linear combination
\algn{\eqnlab{WKBform}
	\hat P(r,s) \sim \frac{A_+}{\sqrt{2\pi{\loc}}} \ete^{-S^+(r,s)/{\loc}} \ete^{-T^+(r,s)} + \frac{A_-}{\sqrt{2\pi{\loc}}} \ete^{-S^-(r,s)/{\loc}} \ete^{ - T^-(r,s)}\,.
}
In \secref{elliptic} we express $S^\pm$ and $T^\pm$ in terms of Carlson elliptic functions~\cite{Car87,Car88} and find that $S^-,T^-\to0$ and $S^+\to\infty$ as $r\to1^-$. Together with \Eqnref{boundcond} this leads to
\algn{\eqnlab{Aval}
	\frac{A_-}{\sqrt{2\pi{\loc}}} = \frac1{2{\abs}}\,.
}
However, taking the limit $r\to 0^+$ shows that the WKB solution diverges and does not fulfil the no-flux condition~\eqnref{noflux}. This problem is remedied by performing asymptotic matching of the WKB solution to the local solution of \Eqnref{lapfp}, valid close to the boundary at $r=0$. As we show in \secref{bl}, the local solution that fulfils the no-flux condition~\eqnref{noflux} is given by
\algn{\eqnlab{hatPbound}
	\hat P(r,s) \sim A_0 \ete^{-\frac{2{\abs} r}{{\loc}}} M\left(\frac{s}{2{\abs\loc}},1,\frac{2{\abs} r}{{\loc}}\right)\,, \quad r\ll 1\,,
}
where $M(a,b,z)$ is a Kummer function~\cite{dlmf}. Approaching the boundary at $r=0$ from the right, $r\to0^+$, we obtain from \Eqnref{hatPbound} the expression
\algn{\eqnlab{P0s}
	\hat P(0^+,s) = A_0(s)\,.
}
The coefficient $A_0(s)$ in turn is obtained by asymptotically matching \Eqnref{hatPbound} with the WKB solution. As we show in \secref{matching}, this gives
\algn{\eqnlab{A0expr}
	A_0(s) = \frac{1}{{\loc}}\Gamma\left(\frac{s}{2{\loc\abs}}\right)\left(\frac{{\loc}}{2{\abs}}\right)^{\frac{s}{2{\loc\abs}}}\,,
}
{where $\Gamma(x)$ denotes the Euler Gamma function~\cite{dlmf}.} In the final step, we use \Eqsref{P0s} and \eqnref{A0expr} together with \Eqnref{logrint}. After Laplace transform, \Eqnref{logrint} reads
\algn{
	\frac{s}{{\loc}} \widehat{\langle \log r\rangle}(s)- \langle\log r\rangle(0)&= {\loc} \hat P(0^+,s)-\frac{2{\abs\loc}}s \,,\nn\\
	&= \Gamma\left(\frac{s}{2{\loc\abs}}\right)\left(\frac{{\loc}}{2{\abs}}\right)^{\frac{s}{2{\loc\abs}}}-\frac{2{\abs\loc}}s\,.
}
Inverse Laplace transform then yields
\algn{
	\dd{\tilde x}\langle \log r\rangle(\tilde x) = \frac1{2\pi i{\loc}}\int_{-i\infty + C}^{i\infty + C}\!\!\ed s\,\ete^{\frac{s \tilde x}{{\loc}}}\left[\Gamma\left(\frac{s}{2{\loc\abs}}\right)\left(\frac{{\loc}}{2{\abs}}\right)^{\frac{s}{2{\loc\abs}}}-\frac{2{\abs\loc}}s\right]\,,
}
\algn{
    = \frac{2{\abs}}{2\pi i}\int_{-i\infty + C}^{i\infty + C}\!\!\ed \hat s\,\ete^{2{\abs}\hat s \tilde x}\left[\Gamma\left(\hat s\right)\ete^{\hat s\log\left(\frac{{\loc}}{2{\abs}}\right)}-\frac{1}{\hat s}\right]\,,
}
where the contour in the complex plane is chosen to stay right of the singularities and we use the rescaled integration variable $\hat s = s/(2{\abs\loc})$.

The integrand has simple poles at $\hat s=-n$ where $n = 1,2,\ldots$. The residues of $\Gamma(\hat s)$ at $\hat s=-n$, $n=1,2,\ldots$ are given by $(-1)^{n}/n!$. Combining this, we replace the integral by a sum over the residues and find
\algn{
	\dd{\tilde x}\langle \log r\rangle(\tilde x) 	&= 		 2{\abs}\sum_{n=1}^\infty\frac{(-1)^n}{n!}\left(\frac{2{\abs}}{{\loc}\ete^{2{\abs} \tilde x}}\right)^n = 2{\abs}\sum_{n=0}^\infty\frac{(-1)^n}{n!}\left(\frac{2{\abs}}{{\loc}\ete^{2{\abs} \tilde x}}\right)^n-2{\abs}\,,\nn\\
						&= 		 2{\abs}\left(\ete^{-\frac{2{\abs}}{{\loc}\ete^{2{\abs}\tilde x}}}-1\right)\,.
}
Comparison with \Eqnref{logrrel} leads to
\algn{\eqnlab{P0val}
	\tilde P(0^+,\tilde x) \sim \frac{2{\abs}}{{\loc}}\ete^{-\frac{2{\abs}}{{\loc}}\ete^{-2{\abs}\tilde x}}\,.
}
This expression shows that the probability $\tilde P(0^+,\tilde x)$ is exponentially small for ${\loc}\ll 1$, and we can read off the WKB contributions by comparison with \Eqnref{wkbssol}
\algn{
	S(0^+,\tilde x) = 2{\abs} \ete^{-2{\abs} \tilde x}\,, \quad	G(0^+,\tilde x) = \sqrt{\frac{2\pi}{{\loc}}}\,.
}
Consistency of the systematic WKB expansion requires that the next order, of order ${\loc}$ in the exponent, be much smaller than the logarithm of the prefactor in \Eqnref{P0val}, i.e., $\log(2{\abs}/{\loc})\ll{\loc}$, from which we find the additional consistency condition ${\rat}\gg1${, i.e., absorption must be much stronger than localization.} Reexpressing \Eqnref{P0val} in the original coordinates gives
\algn{\eqnlab{p0x}
	P(0^+,x) \sim 2{\rat}\ete^{-2{\rat}\ete^{-2 x/{\ell_\text{A}}}}\,.
}
In light of \Eqnref{logrint}, $P(0^+,x)$ is sufficient to obtain $\langle\log r\rangle$ and thus to compute the statistics of resonances in the short-sample limit. We will return to this problem in \Secref{ssl}.

We close this Section on the Laplace-transformed WKB method by noting that due to the approximations made by choosing the Ansatz~\eqnref{wkbssol}, the expression~\eqnref{p0x} predicts an unphysical, non-vanishing probability $P(0^+,0) \sim 2{\rat} \ete^{-2{\rat}}$ at $\tilde x=0$. However, the validity conditions ${\loc}\ll1$ and ${\rat}\gg1$ for the WKB approximation ensure that the error associated with this spurious probability is exponentially small.
\subsection{Resonance density in the semi-infinite sample}\label{lsl}
Now we are in a position to analyze our main object of interest, the density of resonance widths in the complex plane, in the semiclassical regime $\mathcal{E}\gg D^{2/3}$. We first consider the semi-infinite sample limit $L\to \infty$ to showcase the validity of the main relation~\eqref{main_rel}. As has been discussed {in \Secref{meanlog}}, in this limit
 the probability density $P(r,L)=P(r,L)|_{{\cal E},\eta}$ for the reflection coefficient $r$ becomes stationary: $\lim_{L\to\infty}P(r,L) \equiv P_\text{st}(r)$, and is explicitly given by Eq.~\eqref{stationary}. Using this we can immediately evaluate
\be \label{avlog}
	\lim_{L\to \infty}\left\langle \ln{r(L)}\right\rangle =\int_0^{1} \ln(r) P_\text{st}(r)\,\text{d}r=-\ln{(2{\rat})}-\mathbf{C}-e^{2{\rat}}E_1(2{\rat}):=F({\rat}),
\ee
where $\mathbf{C}= 0.57721\ldots$ is Euler's constant and $E_1(a):=\int_a^{\infty}\ete^{-t}\frac{dt}{t}$.
In view of Eq.~\eqref{main_relA}, we restore $\eta$ and ${\cal E}$ in Eq.~\eqref{avlog} via ${\rat}=\frac{4k \eta}{D}=\eta/{\Gamma_\text{L}}$, with  ${\Gamma_\text{L}}=\frac{D}{4\sqrt{{\cal E}}}$ playing the role of the characteristic resonance width at energy ${\cal E}${, see Tab.~\ref{tab:parameters}.}
Differentiating $F({\rat})$ twice and using $k\ell_{L}=4{\cal E}^{3/2}/D$ we obtain
\algn{
	\frac{\partial^2}{\partial \eta^2}F({\rat})&= {\Gamma_\text{L}}^{-2}F''({\rat})\,, \label{derivETA}\\
	\frac{\partial^2}{\partial {\cal E}^2}F({\rat})&= \frac{{\Gamma_\text{L}}^{-2}}{(2k\ell_L)^2}\left[\frac{1}{16}{\rat}^2 F''({\rat})-{\rat} F'({\rat})\right]\,. \label{derivE}
}
As the consistent treatment of the semiclassical regime requires us to only keep the leading order terms in the parameter $k\ell_{L}\gg 1$, we see that  $\frac{\partial^2}{\partial {\cal E}^2}F({\rat})$ is negligible in comparison with $\frac{\partial^2}{\partial \eta^2}F({\rat})$, justifying passage from  Eq.~\eqref{main_relA} to  Eq.~\eqref{main_rel}. Finally, dividing by $4\pi$ yields the density of resonances in the form
\be\label{den_stat}
	\rho^{(\infty)}({\cal E},\Gamma)=\frac{1}{4\pi {\Gamma_\text{L}}^2}F''\left(\frac{\Gamma}{{\Gamma_\text{L}}}\right)\,, \quad \Gamma\ll {\cal E}\,,
\ee
for those resonance poles whose real parts are in the vicinity of the energy value ${\cal E}=k^2$. The condition $\Gamma\ll {\cal E}$, equivalent to ${\mathcal{G}=\Gamma/\Gamma_\text{L}}\ll k\ell_{L}$, ensures that the corrections described by  \eqref{derivE} can be safely neglected. Performing the differentiation of $F({\rat})$ defined in $\eqref{avlog}$  yields the explicit expression for the mean resonance density which can be written in terms of the localization length $\ell_L$  and the wavevector $k$ as 
\be\label{den_statexplicit}
\rho^{(\infty)}_k(\Gamma)	=\frac{\ell^2_L}{kL}\left[\frac{k}{\ell_\text{L}}\frac1{\Gamma}-2\ete^{2\frac{\ell_L}{k}\Gamma}E_1\left(2\frac{\ell_L}{k}\Gamma\right)\right]\,, \quad \Gamma\ll k^2\,,
\ee
equivalent to \eqref{rho_intro_long}. This expression demonstrates the expected behavior $\rho^{(\infty)}_{k}(\Gamma\ll D/k)\sim \ell_\text{L}/(2\pi k \Gamma)\propto\Gamma^{-1}$ which, as we explained in Sec.~\ref{sec:intro}, is related to the exponential localization of eigenfunctions, and has been discussed previously~\cite{Gur12}. The resulting logarithmic divergence of the normalization for $\rho^{(\infty)}_{k}(\Gamma)$ indicates, however, that there should be a cutoff-scale $\Gamma_\text{min}$ below which the resonance density is strongly suppressed.

The size of the cut-off scale  $\Gamma_\text{min}$ can be readily estimated from the following  consideration: The stationary distribution $P_\text{st}(r)$ was found by letting $L\to \infty$. The ensuing resonance density, however, should be valid also for long but finite samples of length $L\gg \ell_\text{L},{\ell_\text{A}}$. The density $\nu(E)$ of energy levels per unit length in such a long sample in the semiclassical limit $k\ell_l\gg 1$ coincides~\cite{Hal65} with  the level density of the free Hamiltonian $H=-\frac{d^2}{dx^2}$ and is given at energy ${\cal E}$ by $\nu(E)=\frac{1}{2\pi}{\cal E}^{-1/2}$. Clearly, this level density must by identical with the density of real parts ${\cal E}_n$ of the resonance positions.  Then our definition \eqref{resdendef} and the relation \eqref{den_stat} imply the relation  
 \be
 L\,\nu({\cal E})\sim\int_{\Gamma_\text{min}}^{\infty}\!\!\rho^{(\infty)}({\cal E},\Gamma)\,\ed\Gamma\sim-\frac{1}{2\pi {\Gamma_\text{L}}}\ln{\frac{\Gamma_\text{min}}{{\Gamma_\text{L}}}}
 \ee
 where we used $\frac{\Gamma_\text{min}}{{\Gamma_\text{L}}}\ll 1$. Solving for $\Gamma$  immediately yields the relation  
 \be\label{cutoff} 
 \frac{\Gamma_\text{min}}{{\Gamma_\text{L}}}\sim \ete^{-L\frac{{\Gamma_\text{L}}}{\sqrt{\cal E}}}=\ete^{-\frac{L}{\ell_L}}=\ete^{-{\loc}}\,,
 \ee
 for $\frac{L}{\ell_L}={\loc}\gg1$. The cut-off scale has a clear physical meaning: the widths of the most narrow resonances in a typical disorder realization  correspond to eigenstates which in the closed sample are localized, due to disorder, in the vicinity of far-left end of the sample. When the sample is open at the right end, these eigenstates are long-lived and require exponentially long times $\sim\ete^{{\loc}}$ to escape to continuum via the open end, translating to exponentially small resonance widths. In the centre part of Tab.~\ref{tab:parameters}, we summarize the different relevant energy scales in the problem, including {$\Gamma_\text{L}$, $\Gamma_\text{S}$,} $\Gamma_\text{min}$, $\mathcal{E}$, and $\eta$.

We thus conclude that the limits $L\to \infty$ and $\Gamma\to 0$ do not commute. Unfortunately, we have not succeeded in finding $P(r,L)|_{{\cal E},\eta}$ for large but finite $L\gg \ell_\text{L},{\ell_\text{A}}$ and finite ${\rat}>0$, which remains a challenging and interesting problem. The lack of such a solution also prevented us from probing the log-normal tail of the distribution, expected to show up at ultranarrow resonances with $\Gamma\ll \Gamma_\text{min}$ \cite{Gur12}.

Coming back to the resonance density~\eqref{den_statexplicit} in the large-sample limit $L\to \infty$, one finds that in the opposite limit of wide resonances $\Gamma\gg D/k$ the tail behavior shows another power-law decay $\rho^{(\infty)}_{k}(\Gamma\gg D/k)\sim (4\pi\Gamma)^{-2}$. Decays of this kind seem to be a universal feature in scattering problems of disordered media for perfect channel coupling. In particular it implies a logarithmic divergence of the mean resonance widths in the semiclassical limit $k\ell_L\to \infty$:
\begin{equation}
	\langle\Gamma\rangle=\frac1{L\,\nu({\cal E})}\int_{\Gamma_\text{min}}^{\cal E}\!\!\rho^{(\infty)}({\cal E},\Gamma) \Gamma\, \text{d}\Gamma\sim \frac{2\sqrt{\mathcal{E}}}{L}\ln{\frac{\cal E}{{\Gamma_\text{L}}}}=\frac{2{\Gamma_\text{L}}}{{\loc}}\ln{(k\ell_L)}\,.
\end{equation}
The existence of this tail regime has been demonstrated recently in  quasi-one-dimensional disordered systems  \cite{Fyo24}, and earlier predicted for the zero-dimensional (quantum chaotic, ergodic) regime, both theoretically, see \cite{Fyo96,Fyo97,Som99,Fyo15} and more recently in \cite{Kur25}, and confirmed experimentally \cite{Che21,Kuh08}.

\begin{figure}
    \includegraphics[]{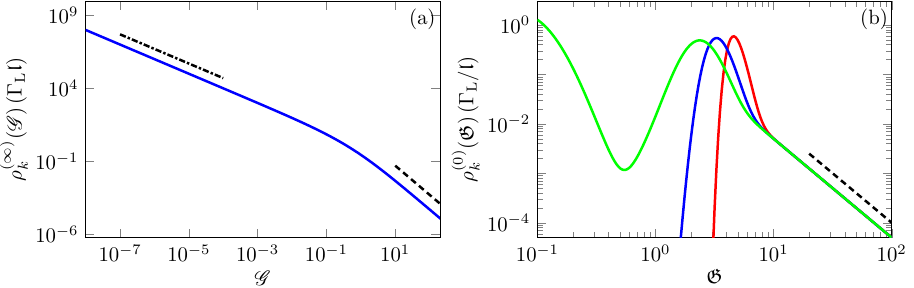}
    \caption{Resonance densities in the limits of short and long samples (solid lines). The dash-dotted and dashed lines show $\Gamma^{-1}$ and $\Gamma^{-2}$ scalings, respectively. (a) $\rho^{(\infty)}_k({\mathcal{G}}) $ as function of ${\mathcal{G}}$ (blue). (b) $\rho^{(\infty)}_k({\mathfrak{G}})$ as function of ${\mathfrak{G}}$ for ${\loc}=10^{-3}$ (red), ${\loc}=10^{-2}$ (blue), and ${\loc}=5\cdot 10^{-2}$ (green).}\label{fig:gamsl}
\end{figure}

Figure~\ref{fig:gamsl}(a) shows the density of resonance widths  $\rho_k^{(\infty)}$ as a function of ${\mathcal{G} = \Gamma/\Gamma_\text{L}}$ obtained from Eq.~\eqref{den_statexplicit}. The small- and large-$\Gamma$ asymptotes are clearly visible.
\subsection{Resonance density in the short-sample limit}\seclab{ssl}
The short-sample corresponds to the case when $\ell_\text{L}\gg L$, so that ${\loc}\ll1$. We also recall 
that the expression~\eqnref{p0x} was obtained under the condition ${\rat}\gg 1$ or, equivalently, $\eta\gg k/\ell_L$. To obtain the density of resonances in this case, we integrate the expression~\eqnref{p0x} over $x$ and use \Eqnref{logrint}, giving
\algn{
		\langle \log r\rangle 	=& 2\left[{\ell_\text{A}}^{-1}\int_0^L\!\!\ed x\,\ete^{-2{\rat}\ete^{-2 x/{\ell_\text{A}}}}-\frac{L}{{\ell_\text{A}}}\right]\,,\\
						=& E_1\left(2{\rat} \text{e}^{-2{\abs}}\right)-E_1(2{\rat}) - 2{\abs}\,.
}
Restoring $\eta$, differentiating twice, and evaluating the result at $\eta=\Gamma$ we obtain the required density of resonance widths as
\algn{\eqnlab{den_short}
	\rho^{(0)}_k(\Gamma) = \frac{{\Gamma_\text{L}}}{2{\loc}\Gamma^2}\left\{\ete^{-2\frac{\ell_\text{L}}{k}\Gamma \ete^{-2\frac{L}{k}\Gamma} }\left[1+2\frac{\ell_\text{L}}{k}\Gamma \ete^{-2\frac{L}{k}\Gamma}\left(1-2\frac{L}{k}\Gamma\right)^2\right]	-\ete^{-2\frac{\ell_\text{L}}{k}\Gamma}\left(1+2\frac{\ell_\text{L}}{k}\Gamma\right)	\right\}\,,
}
valid for wide enough resonances satisfying $\Gamma \gg \Gamma_\text{L}=k/\ell_\text{L}$. In turn, this condition implies $\Gamma/{\cal E}\sim \Gamma/k^2\gg 1/(k\ell_\text{L})$. Hence, in the small disorder limit, where  $1/(k\ell_\text{L})\ll 1$, this range for $\Gamma$ is still compatible with the condition $\Gamma/{\cal E}\ll 1$, allowing one to neglect the second derivative over ${\cal E}$  when evaluating $\rho^{(0)}_k(\Gamma)$. In particular, we see again the universal behavior $\rho^{(0)}_k(\Gamma)\sim(4\pi\Gamma)^{-2}$ in the regime of wide resonances $ k/(2L)\ll\Gamma\ll {\cal E}$ [cf. Eq.~\eqref{den_stat}]. Furthermore, across a narrow region around $\Gamma \sim  \Gamma_\text{S} = k/L$ the resonance density experiences a sharp drop by the exponentially large factor $e^{C/\loc}$, with $C\sim 1$. In the complementary range $ \Gamma_\text{L}\ll\Gamma\ll \Gamma_\text{S}$ the density is further given by
\begin{equation}\label{den_short_A}
	\rho^{(0)}_k(\Gamma) \approx   \frac{1}{{\loc}\Gamma}\, \ete^{-2\frac{\ell_\text{L}}{k}\Gamma }
    \left(\ete^{\frac{4L\ell_\text{L}}{k^2}\Gamma^2}-1\right)=
    \ete^{-2\frac{\Gamma}{\Gamma_\text{L}}} 
     \left\{\begin{array}{cc} \ete^{4\frac{\Gamma^2}{\Gamma_\text{L}\Gamma_\text{S}}}\frac{1}{\mathfrak{l} \Gamma}, &  \sqrt{\Gamma_\text{L}\Gamma_\text{S}}\ll \Gamma\ll \Gamma_\text{S}\,,\\
     4\frac{\Gamma}{\Gamma_\text{L}^{2}} , & \Gamma_\text{L} \ll \Gamma\ll 
     \sqrt{\Gamma_\text{L}\Gamma_\text{S}}\,.\end{array}\right.
  \end{equation}

In particular, this expression shows that before the described drop, $\rho^{(0)}_k(\Gamma)$ decreases for increasing $\Gamma$, before increasing again, and reaching its maximum at around the {typical value $\Gamma\sim \Gamma_\text{S}=   k/L$ for short samples}. However, this decreasing range is exponentially suppressed for small ${\loc}$, and we have not been able to observe it numerically (see Fig.~\figref{rho_gamma} in  \Secref{numerics}).

Figure~\ref{fig:gamsl} shows the density $\rho_k^{(0)}$ of resonance widths as a function of {$\mathfrak{G} = \Gamma/\Gamma_\text{S}= \Gamma L/k$} from \Eqnref{den_short} for different {degrees of localization $\loc$}. The distribution exhibits a sharp peak at ${\mathfrak{G}}\approx1$ and we observe the expected $\Gamma^{-2}$-scaling at large $\Gamma$ in all cases. For small values of ${\loc}$, the decreasing range predicted by Eq.~\eqref{den_short_A} is exponentially suppressed but becomes visible at larger values of ${\loc}\gtrapprox 0.03$. Since the WKB is valid asymptotically for small ${\loc}\ll1$, it is unclear to us if the decreasing behavior described by Eq.~\eqref{den_short_A} is a physical effect or simply an artefact of the approximation. However, as we discuss in \Secref{numerics}, in the bulk of the resonance density the expression~\eqref{den_short_A} agrees excellently with the results of our numerical approaches.

\section{Numerical evaluation of resonance density}\seclab{numerics}
In order to numerically compute the resonances for the system with white-noise potential,  we use four different approaches. The first two are direct methods, similar to those used in Refs.~\cite{Ter00,Her19}, where we discretize the system on a finite grid and obtain the scattering resonances from the (generalized) eigenvalues of the effective Hamiltonian~\eqnref{Heffex}. In addition, we present a third, improved method based on determining the eigenvalues of a slightly different effective Hamiltonian. The fourth method is new and based the main relation~\eqref{main_rel}. We compare these results with our analytical expressions~\eqref{den_stat} and \eqnref{den_short}.
\subsection{Direct numerical methods}
For the direct methods, we discretize the system on a lattice with $N$ sites and lattice spacing $a$, so that $L = N a$, as discussed in Sec.~\ref{sec:intro}.  According to Eq.~\eqref{respoles_eq}, complex resonances are given by the roots of the determinant $\det[z(k) - H_\text{eff}(k)] = 0$, where $z(k) = -2[1-\cos(ka)]/a^2$.
Resonances correspond to solutions of Eq.~\eqref{respoles_eq} whose amplitudes grow along the lead. This condition, and the condition of a right-travelling (outgoing) wave require for $k = u+iv$ that
\algn{\eqnlab{conditions}
	0\leq u\leq \pi\,,\qquad v<0\,.
}
To compare with the continuum model in the semiclassical regime, we need to consider the weak-disorder limit, where the localization length for the discrete model is given by the Thouless expression~\cite{Tho79,Izr12}
\algn{
	\ell^{(l)}_\text{L} = \frac{4\sin^2(k a)}{a^2\xi^2}\,,
}
which reduces to the localization length $\ell_\text{L}$ of the continuum model in \Eqnref{ellloc} for $ka\ll1$. 

The behavior of the system depends only on the ratio between the localization length $\ell_\text{L}$ and length $L$ of the medium, so for the continuous system the ratio ${\loc} = L/\ell_\text{L}$ represents the characteristic quantity. The corresponding quantity for the discrete lattice model reads ${\loc}^{(l)} = N/\ell^{(l)}_\text{L} = Na^2\xi^2/[4\sin^2(ka)]$. In order to lighten the notation, we will drop the superscript $(l)$ in what follows, whenever the distinction between lattice and continuum descriptions is either unambiguous or irrelevant.
\subsubsection{Exact resonances}\seclab{exact}
Exact resonances are obtained by numerically finding the solutions of Eq.~\eqref{respoles_eq} in the $u$-$v$ plane. We compute the exact resonances using two different strategies. First, following Ref.~\cite{Ter00}, we numerically compute variation of the complex phase $\Delta\phi$ of $\det[z(k) - H_\text{eff}(k)]$ along closed loops of $z(k)$ in the complex plane. Along a closed loop, $\Delta\phi$ is given by
\algn{\eqnlab{phasechange}
	\Delta\phi = 2\pi N_\text{res}\,,
}
where $N_\text{res}$ denotes the number of resonances, i.e., the number of zeros of $\det[z - H_\text{eff}(z)]$ enclosed by the loop. This method is convenient for the accurate computation of the resonance distribution. However, it does not resolve individual resonances. To achieve this, we employ a root finding scheme based on the Newton-Raphson method to determine the complex roots of $\det[z - H_\text{eff}(z)]$. 
\subsubsection{Parametric resonances}\seclab{parametric}
Parametric resonances~\cite{Vin12,Her19} are obtained by directly computing the eigenvalues of the effective Hamiltonian $H_\text{eff}$ in \Eqnref{Heffex} around $k a=\pi/2$, where the density is insensitive to changes of the energy $E$. To analyse parametric resonances, we define the centred energies $\tilde E := E-2/a^2 = -2\cos( ka )/a^2$ and the corresponding effective Hamiltonian $\tilde H_\text{eff}(\tilde E) := H_\text{eff}(\tilde E) - 2/a^2 {\bf 1}_N$. Using \Eqnref{Heffex}, $\tilde H_\text{eff}(\tilde E)$ can be written as
\algn{
	\tilde H_\text{eff}(\tilde E) = H_\text{in} - \frac{2}{a^2} {\bf 1}_N +t^2\left(\frac{\tilde E}2 - i \sqrt{\frac1{a^{4}}-\frac{\tilde E^2}4 }\right)\,|N\rangle \langle N|
}
For $k a \approx \pi/2$, we have $|\tilde E|\ll 1$, which justifies the Taylor expansion
\algn{\eqnlab{Heffexp}
	\tilde H_\text{eff}(\tilde E) \sim \tilde H_\text{eff}(0) + \tilde H'_\text{eff}(0)\tilde E\,.
}
Keeping only the zeroth order term, $\tilde H_\text{eff}$ becomes independent of $\tilde E$ and reads
\algn{\eqnlab{Heffpar}
	\tilde H_\text{eff} \approx \tilde H_\text{eff}(0) =  H_\text{in}- \frac{2}{a^2} {\bf 1}_N -i\frac{t^2}{a^2}\,|N\rangle \langle N|\,.
}
The so-called parametric resonances correspond to the complex eigenvalues of the $\tilde E$-independent effective Hamiltonian \eqnref{Heffpar}. They are simpler to compute because the corresponding eigenvalue problem
\algn{\eqnlab{Heffpareig}
	\tilde H_\text{eff}(0)\psi =  \tilde E_\text{par}\psi\,,
}
is regular. Parametric resonances should agree with the exact eigenvalues for small $\tilde E$, or whenever the densities are insensitive to changes of $\tilde E$. This is the case for weak disorder close to centre of the spectrum at $ka \approx \pi/2$, as we shall see.

An improved approximation, which is new to the best of our knowledge, is obtained by retaining the linear term in \Eqnref{Heffexp}. In this case, the resulting equation can again be recast in a regular eigenvalue problem which now reads
\algn{\eqnlab{Heffconeig}
	\left[{\bf 1}_N - \tilde H_\text{eff}'(0)\right]^{-1} \tilde H_\text{eff}(0)\psi =  \tilde E^*_\text{par}\psi\,,
}
where $\tilde H_\text{eff}'(0) = (t^2/2)|N\rangle \langle N|$. This approximation has the advantage that it consistently keeps $\tilde E$ up to first order, which turns out to significantly increase its accuracy.
\subsubsection{Comparison of different numerical schemes}
\begin{figure}
	\includegraphics[width = \linewidth]{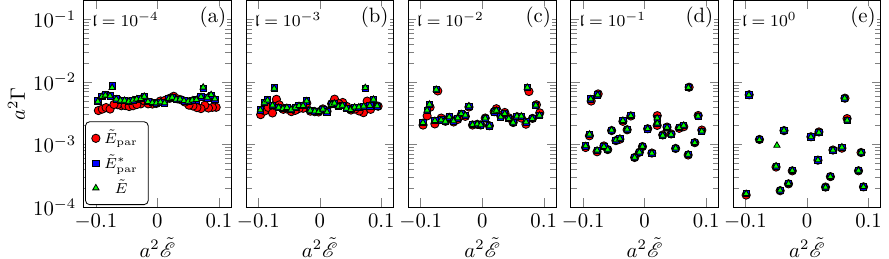}
	\caption{Comparison between exact resonances $\tilde E$ and the two different kinds of parametric resonances $\tilde E_\text{par}$ and $\tilde E^*_\text{par}$ described in the main text, for different ${\loc}$ [(a)--(e)] obtained from a single realization of the disorder with $N=10^3$, $a=t=1$.}\figlab{comparison}
\end{figure}
Figure~\figref{comparison} shows a sample of exact resonances together with the two approximated types of parametric resonances for a single realization {of the disorder} and for different localization lengths. We find that for {strong enough localization, i.e.,} ${\loc}= L/\ell_\text{L}>0.1$, both approximations agree well with the exact resonances. For smaller ${\loc}$, in the ballistic regime, the resonances widths have larger values and the standard parametric approximation based on \Eqnref{Heffpar} visibly deviates from the exact resonances. The improved approximation is visibly more accurate.

\begin{figure}
	\includegraphics[width = \linewidth]{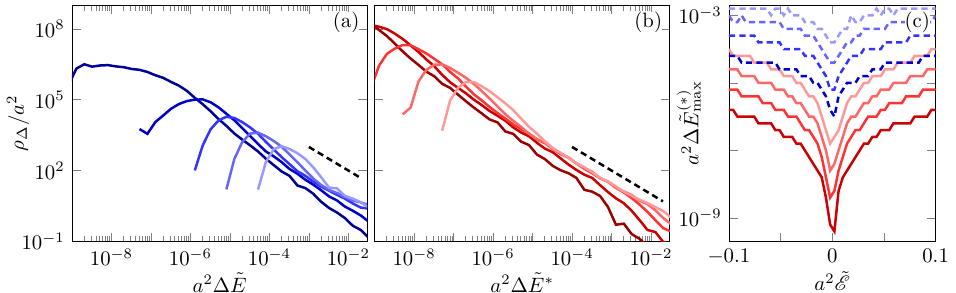}
	\caption{Distribution of differences $\Delta \tilde E^{(*)}$ between exact and parametric energy values for $N=10^3$ and $t=a=1$, obtained from $10^4$ realizations. (a) Distribution $\rho_\Delta$ of differences between exact and parametric resonances based on \Eqnref{Heffpareig} for ${\loc}=2.5\cdot 10^{-4}$, ${\loc}=2.5\cdot 10^{-3}$, ${\loc}=2.5\cdot 10^{-2}$, ${\loc}=0.25$, and ${\loc}=2.5$ in blue colours from light to dark. (b) Same as in (a) but for differences between exact and improved parametric resonances based on \Eqnref{Heffconeig}. The ${\loc}$ values are the same as in (a) but shown in red colours from light to dark. (c) Locations ${\Delta \tilde E^{(*)}}_\text{max}$ of the maxima of the distributions in (a) [blue, dashed] and (b) [red, solid] as functions of the energy $\mathcal{E}$ for ${\loc}=2.5\cdot 10^{-4}$, ${\loc}=2.5\cdot 10^{-3}$, ${\loc}=2.5\cdot 10^{-2}$, and ${\loc}=0.25$ from light to dark colours.}\figlab{distcomparison}
\end{figure}

This is confirmed by the results in \Figref{distcomparison}: Figures~\figref{distcomparison}(a) and \figref{distcomparison}(b) show the distributions $\rho_{\Delta}$ of differences $\Delta \tilde E^{(*)} = |\tilde E - \tilde E^{(*)}|$ for $a^2\mathcal{E}\ll 1$ between exact and parametric resonances for different ${\loc}$. As ${\loc}$ increases, all differences become smaller, as observed in \Figref{comparison}. However, the differences for the improved parametric resonances $\Delta \tilde E^{*}$ in \Figref{distcomparison}(b) are typically several orders of magnitude smaller than those of the regular parametric resonances $\Delta \tilde E$ in \Figref{distcomparison}(a). In all cases the distributions have a maximum at small differences, but show a power-law tail $\sim {\Delta \tilde E^{(*)}}^{-1}$ at large distances. Figure~\figref{distcomparison}(c) shows the location of the maximum $\Delta \tilde E^{(*)}_\text{max}$ as a function of the energy $\mathcal{\tilde E}$ for different ${\loc}$. We again observe that ${\Delta \tilde E^{*}}_\text{max}$ is consistently smaller than ${\Delta \tilde E}_\text{max}$ and that all quantities are smallest close to the $\mathcal{\tilde E}=0$. This is consistent with the expectation that the approximation using parametric resonances should be most accurate close to the centre of the spectrum. Upon increasing the system size $N\to\infty$ we find that all distances decrease around $\mathcal{\tilde E}=0$ (not shown).

\begin{figure}
	\includegraphics[width=\linewidth]{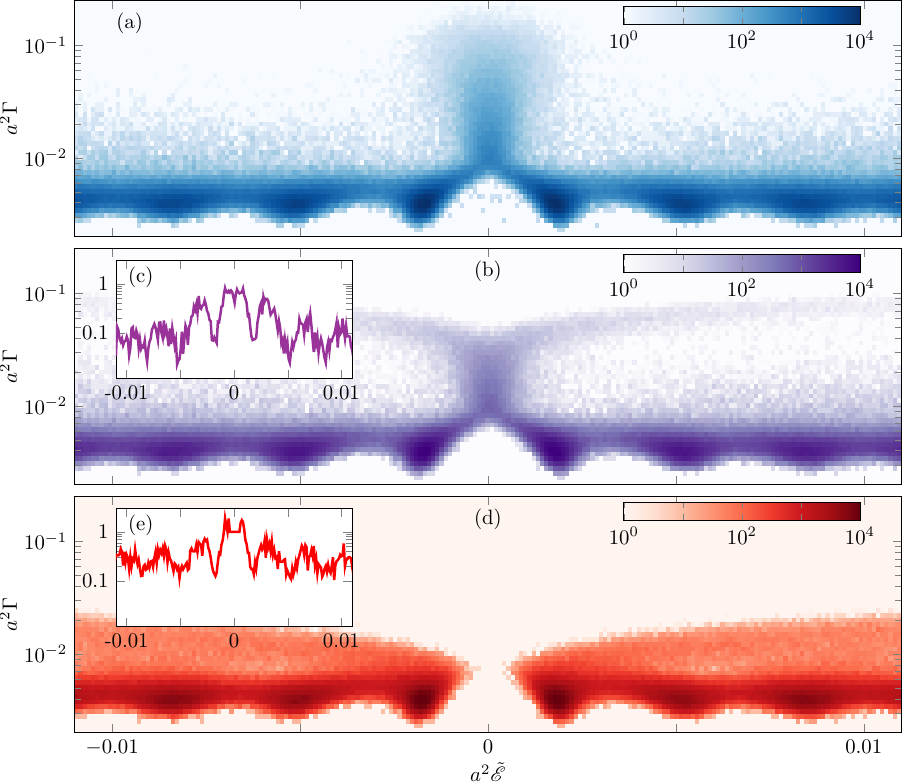}
	\caption{(a) Numerically obtained resonance densities in the complex plane as functions of $\mathcal{\tilde E}$ and $\Gamma$ for $N=10^3$, ${\loc} = 2.5\cdot10^{-3}$ from $10^5$ realizations. (a) Density of exact resonances obtained from numerically finding the roots of the determinant. (b) Densities of improved parametric resonances obtained from numerically solving the eigenvalue problem in \Eqnref{Heffconeig}. {(c) $L^1$ distance between the resonance densities in (a) and (b) as a function of $\mathscr{\tilde E}$.} (d) Density of parametric resonances obtained from numerically solving the eigenvalue problem in \Eqnref{Heffpareig}. {(e) $L^1$ distance between the resonance densities in (a) and (d) as a function of $\mathscr{\tilde E}$.}}\figlab{engam}
\end{figure}

The differences between exact and parametric resonances can also be observed on the level of the densities of resonance widths. To highlight the differences, we focus on the narrow region around $\mathcal{\tilde E}\approx0$ where both approximation schemes are expected to be quite accurate.
Figure~\figref{engam} shows {the} numerically computed densities $\rho(\Gamma,\mathcal{\tilde E})$ for $\mathcal{\tilde E}\approx0$ in the ballistic regime from root finding [\Figref{engam}(a)], from \Eqnref{Heffconeig}  [\Figref{engam}(b)], and from \Eqnref{Heffpareig}  [\Figref{engam}(d)]. We observe significant differences between the densities, in particular for large $\Gamma$ and in a very small region around $\mathcal{\tilde E}=0$. {This is reflected in the $L^1$ distances between the exact and (improved) parametric resonance densities shown as functions of $\mathcal{\tilde E}$ in \Figsref{engam}(c) and \figref{engam}(e). These distances reach values of the order of unity very close to $\mathcal{\tilde E}=0$.}

 Outside the ballistic regime, the picture is {notably} different: Figure~\figref{engam2} shows the same densities, but for ${\loc}=0.25$. In this case, the densities in \Figsref{engam2}(a), {\figref{engam2}(b), and \figref{engam2}(d)} are practically indistinguishable. {Consequently, the $L^1$ distances between the exact and (improved) parametric resonance densities, shown in \Figsref{engam2}(c) and \figref{engam2}(e), are substantially smaller than in \Figref{engam}, of the order of $0.1$. In both \Figsref{engam} and \figref{engam2}, i.e., both within and outside the ballistic regime, the densities of improved parametric resonances are consistently closer to the densities of exact resonances, cf. \Figsref{engam}(c) and \figref{engam}(e), as well as \Figsref{engam2}(c) and \figref{engam2}(e).}

\begin{figure}
	\includegraphics[width=\linewidth]{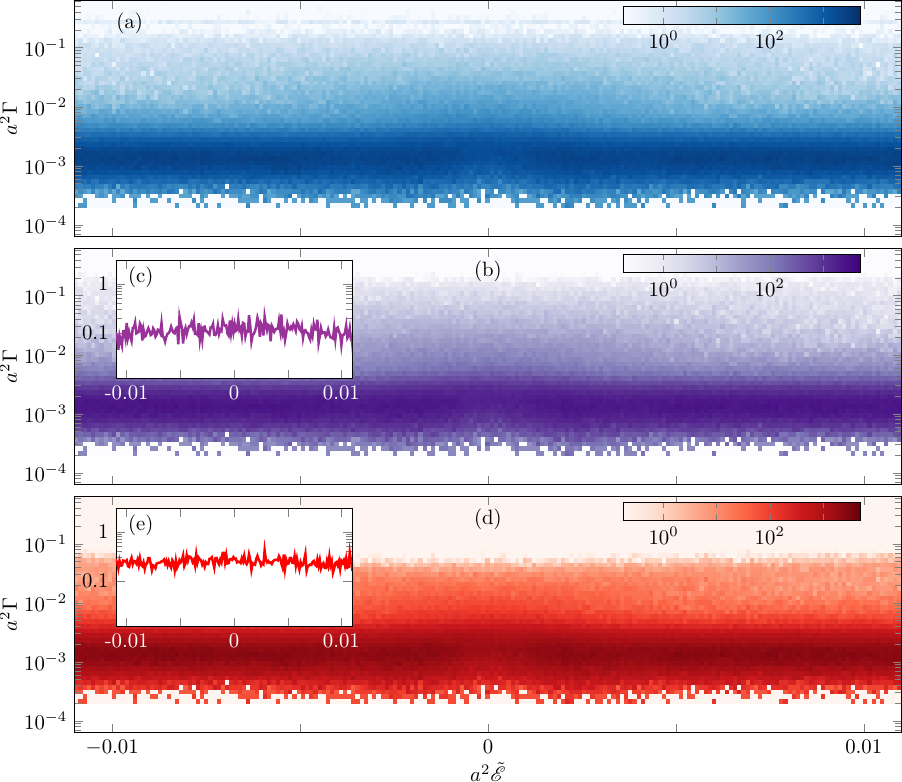}
	\caption{(a) Numerically obtained resonance densities in the complex plane as functions of $\mathcal{\tilde E}$ and $\Gamma$ for $N=10^3$, ${\loc} = 2.5\cdot10^{-1}$ from $10^5$ realizations. (a) Density of exact resonances obtained from numerically finding the roots of the determinant. (b) Densities of improved parametric resonances obtained from numerically solving the eigenvalue problem in \Eqnref{Heffconeig}. {(c) $L^1$ distance between the resonance densities in (a) and (b) as a function of $\mathscr{\tilde E}$.} (d) Density of parametric resonances obtained from numerically solving the eigenvalue problem in \Eqnref{Heffpareig}. {(e) $L^1$ distance between the resonance densities in (a) and (d) as a function of $\mathscr{\tilde E}$.}}\figlab{engam2}
\end{figure}
We conclude that parametric resonances represent useful approximations for the exact resonances at large system size. The improved parametric resonances based on \Eqnref{Heffconeig} are consistent to first order in $\tilde E$ and are significantly more accurate than the standard parametric resonances based on \Eqnref{Heffpareig}, whilst retaining the computational advantages of dealing with a regular eigenvalue problem.
\subsubsection{Comparison with theory}
\begin{figure}
	\includegraphics[width = \linewidth]{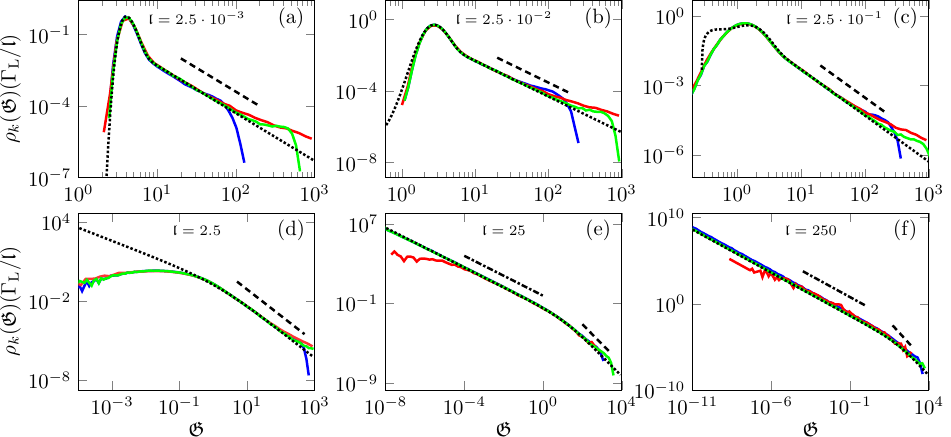}
	\caption{Density $\rho_k({\mathfrak{G}})$ of scattering resonances for different ${\loc} = L/\ell_\text{L}$, from theory (dotted lines) and from numerical computations of exact resonances (red), standard parametric resonances based on \Eqnref{Heffpareig} (blue), and improved resonances based on \Eqnref{Heffconeig} (green). Parameters: $N=5\cdot 10^3$, $t=a=1$. The dashed and dash-dotted lines show $\Gamma^{-2}$ and $\Gamma^{-1}$-scaling, respectively. The top row explores the range ${\loc}\ll 1$, when theoretical curves are given by \eqnref{den_short}, while the bottom row explores the range ${\loc}\gg 1$, with theoretical curves described by \eqref{den_statexplicit}.}\figlab{rho_gamma}
\end{figure}
We now compare the numerical data for the density of resonances of the discrete model with the theory developed above. To this end, the numerically obtained density of resonance widths obtained for the discrete model with lattice spacing $a=1$ must be rescaled in a way described in \secref{numtheo} to comply with the results of the continuum model. We recall that ${\rat} = \ell_\text{L} \eta/k$ and ${\abs} = L \eta/k${, as well as $\mathcal{G} = \rat|_{\eta=\Gamma} = \Gamma/\Gamma_\text{L}$ and $\mathfrak{G} = \abs|_{\eta=\Gamma}=\Gamma/\Gamma_\text{S}$}, see Tab.~\ref{tab:parameters}.

Figure~\figref{rho_gamma} shows the density $\rho_k$ of resonance widths as a function of ${\mathfrak{G}=\Gamma/\Gamma_\text{S}}$, for different values of ${\loc}$ from numerics and theory. Figures~\figref{rho_gamma}(a)--(c) show the resonance density in the ballistic regime, where the samples are short compared to the localization length, i.e., {localization is weak,} ${\loc} = L/\ell_\text{L}\ll1$. For the shortest samples shown in \Figref{rho_gamma}(a) and (b) and for large enough ${\mathfrak{G}}$, the numerically computed densities from the discrete model are in good agreement with the analytic expression [\Eqnref{den_short}], obtained from the Laplace-transformed WKB method presented in \Secref{ltwkb}. All curves show a prominent $\sim\Gamma^{-2}$ scaling (dashed line) at large $\Gamma$, which is, however, cut off prematurely for the standard parametric resonances based on \Eqnref{Heffpareig}. For ${\loc}=0.25$, shown in \Figref{rho_gamma}(c), the analytic resonance density obtained from \Eqnref{den_short} shows noticeable deviations from the numerical data, indicating the breakdown of the WKB approximation. Figures~\figref{rho_gamma}(d)--(f) show how the resonance density approaches the limit of long sample size $L\to\infty$, i.e., the localized regime with ${\loc}\gg1$. In this limit, the resonance density is governed by the explicit expression~\eqref{den_statexplicit}, obtained from the stationary distribution of the reflection coefficient. Figures~\figref{rho_gamma}(e) and \figref{rho_gamma}(f) show good agreement between theory and numerics. Notably, all approaches show both the small $\Gamma^{-1}$ (dash-dotted) and the large $\Gamma^{-2}$ scaling. For {moderate localization with $\loc$} of the order of unity [\Figref{rho_gamma}(d)] the system is in an intermediate regime, where~\eqref{den_statexplicit} breaks down.

In conclusion, the numerically obtained resonance density from the discrete model is in good agreement with the results of the theory for the ballistic and the localized regime. However, both theories are valid asymptotically for ${\loc}\ll1$ and ${\loc}\gg1$ and break down in an intermediate regime at ${\loc}$ values of order unity. 
\subsection{Numerical methods based on main relation}\seclab{num_main_rel}
Finally, we point out the possibility to use the main relation~\eqref{main_rel} for numerical simulations in all regimes, including the intermediate regime where our analytical results are not valid. This can be done by computing $\langle \log r\rangle$ numerically for a range of $\eta$ values, either from simulated trajectories using Eq.~\eqnref{rIto}, from numerical evaluation of the Fokker-Planck equation, or from numerically evaluated approximation schemes. 

Figure~\figref{gam_numerical}(a) shows negative $\langle\log r\rangle$ as a function of ${\abs}$, computed numerically from the Langevin equation~\eqnref{rIto} (blue colours) and from the real-space WKB approximation (red colours) for different ${\loc}$. We observe that the general behavior of $-\langle\log r\rangle$ is the same for both methods, although the result from the WKB approximation visibly deviates from that of the Langevin equation as ${\loc}$ increases. For small ${\abs}$ and small ${\loc}$, $\langle\log r\rangle$ behaves as $\langle\log r\rangle \sim -2{\abs}$ (black dashed line). This case corresponds to probability distributions $P(r,L)$ of the reflection coefficients [see \Figref{distributions}(a) and \figref{distributions}(b)] that hardly interact with the boundary at $r=0$ at all. For larger ${\abs}$, by contrast,  $P(r,L)$ strongly interacts with this boundary. As a consequence, $-\langle\log r\rangle$ asymptotically increases more slowly, i.e., logarithmically (dotted line), for large ${\abs}$. After obtaining the resonance density through \eqref{main_rel} by taking the second derivative {and evaluating at $\eta=\Gamma$}, this logarithmic growth leads to the $\Gamma^{-2}$ tail of the resonance densities in \Figsref{distributions}. This is consistent with the intuition that long-lived resonances are associated with a small reflection coefficient $r\ll 1$, i.e., they decay slowly. Therefore, the $\Gamma^{-2}$ decay of the resonance density is intimately related with $P(r,L)$ that strongly interact with the boundary at $r=0$.

From the numerically obtained $\langle \log r\rangle$ we evaluate the second derivative with respect to $\eta$ to obtain the resonance density through the main relation \eqref{main_rel}. The standard discretization of the discrete second derivative through the three-point formula is, however, noisy for real data. This can be mitigated by using filtering techniques like the Savitzky-Golay filter, which uses local polynomial fits of the data over a finite window of data points. After filtering, the derivatives are applied to these polynomials, which typically leads to a smoother result.

\begin{figure}
	\includegraphics[width=\linewidth]{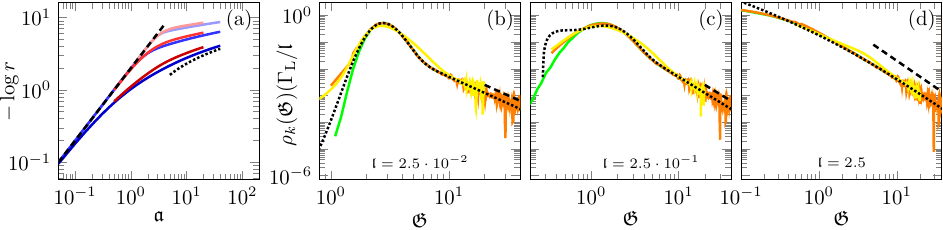}
	\caption{Numerically obtained $-\langle\log r\rangle$ and resonance densities of resonance widths $\rho_k({\mathfrak{G}})$ for different ${\loc}$. The averages from the Langevin simulations are obtained over $10^8$ realizations. (a) Negative $\langle \log r\rangle$ as function of ${\abs}$ obtained from Langevin simulations (blue) and from the numerical WKB method (red) for different ${\loc}=0.025$, $0.25$, and $2.5$ (from light to dark colours). (b)--(d) Resonance densities shown as functions of ${\mathfrak{G}=\Gamma/\Gamma_\text{S}}$, obtained from taking filtered ${\abs}$ derivatives of the data shown in panel (a). The orange lines are from the data of the Langevin simulations, the yellow lines correspond to the numerical WKB approximation. The green and dotted lines are the same as those in \Figsref{rho_gamma}(b)--(d) and are shown for comparison. The black dashed line shows $\Gamma^{-2}$ scaling.}
	\figlab{gam_numerical}
\end{figure}

Figures~\figref{gam_numerical}(b)--(d) show the resonance densities as functions of ${\mathfrak{G}}$ obtained by taking filtered second derivatives of $\langle\log r\rangle$ from Langevin equations (orange) and from the numerical WKB-approximation (yellow). The green and black dashed lines are the same as in \Figsref{rho_gamma}(b)--(d), obtained from evaluation of \Eqnref{Heffconeig} for the discrete model and from theory, respectively. We observe excellent agreement between the resonance density obtained through the Langevin equation and those of the discrete model around the maximum of the distribution. However, we observe deviances for small ${\mathfrak{G}}$ which are likely due to discretization errors when numerically solving the Langevin equation. The numerical WKB-approximation shows deviations from the other methods, which are likely due to the fact that the WKB approximation is accurate asymptotically for ${\loc}\ll1$.

In summary, this brief analysis shows that the main relation~\eqref{main_rel} gives rise to new ways to compute the resonance density numerically. Without further optimization, however, the brute-force method of numerically taking ${\abs}$ derivatives of $\langle \log r\rangle$ {and evaluating at $\eta=\Gamma$}, as done here, is more challenging that obtaining the complex eigenvalues of the discretized problem, through \Eqsref{Heffpareig} or \eqnref{Heffconeig}.

\section{Conclusions}
In this work, we have developed an analytic approach to evaluating the density $\rho ({\cal E},\Gamma)$ of complex resonance poles in the one-dimensional pure reflection off a one-dimensional disordered medium. Building upon the ideas of Refs.~\cite{Che21,Fyo24}, we have established a link connecting $\rho ({\cal E},\Gamma)$ to the distribution $P(r,L)|_{{\cal E},\eta}$ of the reflection coefficient $r$ at {\it complex} energies $E = {\cal E} +i\eta$, through the main relation~\eqref{main_rel}. Physically, the parameter $\eta>0$ is associated with the uniform rate of absorption within the medium. In the case of white-noise disorder, the distribution $P(r,L)|_{{\cal E},\eta}$ satisfies a Fokker-Planck equation~\cite{Pra94,Fre97,Mis97}, which can be solved and approximated with established methods~\cite{Ris89}.

Leveraging~\eqref{main_rel} allows for the analysis of the problem in new parameter regimes. Namely, in the limits of a (semi)infinite and a short medium (compared to the localization length), we obtain the explicit formulas~\eqref{den_statexplicit} and \eqnref{den_short}, which go considerably beyond known results. This way, the theory describes a crossover from narrow to broad resonances and grants analytical access to parameter regimes that have not been systematically addressed before. In the (semi)infinite sample limit the resonance density has been related to the stationary distribution of the reflection coefficient which is obtained in an explicit analytic form. In the short-sample (ballistic) limit, which has not been systematically studied before, we have devised a WKB approximation, both of the original Fokker-Planck equation and of its Laplace transform. The former, direct approach provides a new way to compute the density of resonance widths numerically for short samples. The latter WKB approximation of the Laplace-transformed Fokker-Planck equation led us to an asymptotic matching problem, which we have been able to solve analytically in the short-sample limit, leading to \Eqnref{den_short}.

Furthermore, we have studied the statistics of resonance widths numerically for the discrete version of the model. Here, we have introduced an internally consistent approximation for scattering resonances through the eigenvalue problem~\eqnref{Heffconeig}, which we showed to be significantly more accurate than the standard parametric approximation \eqnref{Heffpareig}. Comparing the results from numerics and theory, we observed a good agreement. Finally, we noted that the main relation~\eqref{main_rel} in principle allows for new numerical approaches to compute the density of resonance widths, but requires further optimization, which goes beyond the scope of this work.

Unfortunately, we were unable to obtain an expression for the density of ultralong-lived resonances for both short and long sample lengths. In this regime, the resonance density has been argued to show a log-normal behavior in Ref.~\cite{Gur12}, but both the stationary state and our WKB approximations break down here. We therefore leave this as a future problem, for which the main relation~\eqref{main_rel} could still be a useful tool.

It would also be interesting to extend the present approach to quasi one-dimensional media, which are described by a system of a few transversely coupled one-dimensional systems. In this case, we expect the main relation~\eqref{main_rel} to remain valid, with the role of $r$ played by squared modulus of the determinant of the reflection matrix, and expect this problem requiring to solve higher-dimensional Fokker-Planck equations.  While the short-medium (WKB) approximation for such systems seems challenging, the case of long samples may be analytically accessible via relating to the methods of Refs.~\cite{Bru96,Mis97}. This approach may help to bridge between the statistics of resonances in the present, strictly one-dimensional situation and the one obtained in Ref.~\cite{Fyo24} in the nonlinear sigma-model approach to disordered wires. The latter model is expected to faithfully describe the limit of an asymptotically large number of strongly coupled one-dimensional systems~\cite{Bro96}.

Finally, controlled approximations of the distribution of resonance widths in various regimes of disordered systems of higher dimensions, in particular in the vicinity of the Anderson localization transition, remain a largely open problems, although some results in this direction have been obtained~\cite{Kot02,Pin04,Wei06,Ski06,Men06,Ski11,Ski16,Gas22,Mar23,Fyo24}. 

We leave the analysis of these intriguing questions to future studies.  
\section{Acknowledgements}
This work was initiated during JM's stay at King's College London as Feodor Lynen Research Fellow supported by Alexander von Humboldt-Foundation. YVF's research at King's College London was supported by  EPSRC grant UKRI1015 ``Non-Hermitian random matrices: theory and applications''.
\appendix
\section{Connection between numerics and theory}\seclab{numtheo}
The {density} of resonance widths $\Gamma$ at a given (centred) energy ${\cal \tilde E}=\mathcal{E}-2/a^2$ is given by
\algn{
	\rho_\mathcal{\tilde E}(\Gamma) = \frac{1}{N\nu({\cal \tilde E})}\left\langle\sum_{n=1}^N \delta({\cal \tilde E}-{\cal E}_n)\delta(\Gamma-\Gamma_n)\right\rangle\,,
}
where $\nu({\cal \tilde E})$ denotes the mean energy level density at weak disorder, normalized as
\algn{
	\nu({\cal \tilde E}) = \frac{1}{N} \left\langle	\sum_{n=1}^N \delta({\cal \tilde E}-{\cal E}_n)\right\rangle \approx \frac{a^2}{2\pi}\frac1{\sqrt{1-\left(\frac{{\cal \tilde E}a^2}{2}\right)^2}}\,,
}
implying that $\rho_{\cal \tilde E}(\Gamma)$ is normalized to unity. To compare with the continuum limit $a\to0$ we need to evaluate $\nu({\cal \tilde E})$ in the bulk (centre) of the distribution, ${\cal E}\approx0$. We then have
\algn{
	\nu({\cal E}\approx0) \approx \frac{a^2}{2\pi}\,.
}
The main relation~\eqref{main_rel} reads
\algn{
	\left\langle\sum_{n=1}^N \delta({\cal \tilde E}-{\cal \tilde E}_n)\delta(\Gamma-\Gamma_n)\right\rangle = \frac1{4\pi}\frac{\partial^2}{\partial\Gamma^2}\left\langle\ln r({\cal \tilde E},\eta = \Gamma)\right\rangle\,.
}
In the continuous description for large samples $L\gg\ell_\text{L}$, we make use of the parameters 
\algn{
	{\rat} = \frac{4k}{D}\eta\,, \quad 	{\mathcal{G}} = \frac{4k}{D}\Gamma\,.
}
Rewriting the main relation in terms of ${\rat}$ and ${\mathcal{G}}$ gives
\algn{
	\left\langle\sum_{n=1}^N \delta({\cal \tilde E}-{\cal \tilde E}_n)\delta(\Gamma-\Gamma_n)\right\rangle = \left(\frac{4k}{D}\right)^2\frac1{4\pi}\frac{\partial^2}{\partial{\rat}^2}\left\langle\ln r({\cal \tilde E},{\rat})\right\rangle|_{{\rat=\mathcal{G}}} = \Gamma_\text{L}^{-2}\rho({\cal \tilde E},{\mathcal{G}})\,,
}
implying further
\algn{
	\rho_ {\cal \tilde E}(\Gamma) = \frac1{N\nu({\cal E})}\Gamma_\text{L}^{-2}\rho({\cal \tilde E},{\mathcal{G}})\,,
}
and, hence,
\algn{
	\rho({\cal \tilde E},{\mathcal{G}}) = N\nu({\cal{\tilde E}}\approx 0)\Gamma_\text{L}^{2} \rho_{\cal \tilde E}\left(\Gamma_\text{L}{\mathcal{G}}\right) = \frac{a^2N}{2\pi}\Gamma_\text{L}^{2} \rho_{\cal \tilde E}\left(\Gamma_\text{L}{\mathcal{G}}\right)\,.
}
For the variables characterizing short samples, we have
\algn{
	{\abs} = \frac{L}{k}\eta = \frac{N a}{k}\eta\,, \quad {\mathfrak{G}} = \frac{L}{k}\Gamma = \frac{N a}{k}\Gamma\,.
}
In terms of these variables, analogous considerations lead us to
\algn{
	\rho({\cal \tilde E},{\mathfrak{G}}) = \frac{a^2N}{2\pi}\Gamma_\text{S}^2 \rho_{\cal \tilde E}\left(\Gamma_\text{S}{\mathfrak{G}}\right) = \frac{1}{2\pi}\frac{k^2}{N} \rho_{\cal \tilde E}\left(\Gamma_\text{S}{\mathfrak{G}}\right)\,.
}
In order to compare with the lattice numerics, we have to match the localization lenghts of both cases. We have
\algn{\eqnlab{identification}
	\frac{4 a \sin^2(k^{(l)} a)}{\xi^2} = \frac{4 {k^{(c)}}^2}{D}\,,
}
where the left- and right-hand sides constitute the localization lenght on the lattice and in the continuum, respectively. Accordingly, $k^{(l)}$ and $k^{(c)}$ denote the lattice and continuum wave numbers. We want to evaluate the resonances in the bulk of energy spectrum which is why we need to evaluate
 the lattice wave number $k^{(l)}$ at $k^{(l)}\approx\frac{\pi}{2a}$. We then make the following identifications for the continuum system
\algn{
	k^{(c)} = \frac1a\,, \qquad D= \xi^2\,,
}
in order to satisfy the condition~\eqnref{identification}. Finally, we match the lattice numerics with the continuum results by making the identifications
\algn{
	\rho({\cal \tilde E},{\mathcal{G}}) =& \frac{a^2N}{2\pi}\left(\frac{\xi^2a}4\right)^2 \rho_{\cal \tilde E}\left(\frac{\xi^2 a}{4}{\mathcal{G}}\right)\,,\\
	\rho({\cal \tilde E},{\mathfrak{G}}) =& \frac{1}{2\pi}\frac1{Na^2} \rho_{\cal \tilde E}\left(\frac{k}{N a}{\mathfrak{G}}\right)\,,
}
where the right-hand sides are the densities of resonance width obtained numerically as described in Sec.~\secref{numerics}.
\section{Solution of Laplace-transformed WKB in terms of elliptic functions}\seclab{elliptic}
The Laplace-transformed WKB Ansatz in \Eqnref{WKBform} depends on the two solutions $S^\pm(r,s) \sim S^\pm(r,s) +{\loc} T^\pm(r,s)$ of the system of equations~\eqnref{steqs}. We first express $S(r,s)$ as an integral
\algn{\eqnlab{S0pmint}
	S^\pm(r,s) = {\abs}\int_{r^\pm_0}^r\ed q\frac{q\pm\sqrt{q \left[q+\frac{s}{{\abs}^2}(1-q)^2 \right]}}{q(1-q)^2} = {\abs}\int_{r_0^\pm}^r\!\!\!\ed q \frac{q\pm \sqrt{Q(q)}}{q(1-q)^2}\,,
}
where we defined
\algn{
	Q(q) \equiv q \left[q+\frac{s}{{\abs}^2}(1-q)^2 \right] = \frac{s}{{\abs}^2} q (q-q_+)(q-q_-)\,,
}
with
\algn{
	q_\pm = 1-\frac{{\abs}^2}{2s} \pm \sqrt{\left(1-\frac{{\abs}^2}{2s}\right)^2 - 1}\,.
}
The constants $r_0^\pm$ are free parameters that can be chosen by convenience, as they will only affect the prefactors in \Eqnref{WKBform}. 

We transform the integral in \Eqnref{S0pmint} into a combination of Carlson elliptic integrals of the first and second kind,
\algn{
	R_F(x,y,z) =& \frac12 \int_0^\infty\!\!\!\ed u\,\frac1{\sqrt{u+x}\sqrt{u+y}\sqrt{u+z}}\,,\\
	R_D(x,y,z) =& \frac32\int_0^\infty\!\!\!\ed u\,\frac1{\sqrt{(u+x)(u+y)(u+z)}(u+z)}\,,
}
respectively, using the identities (2.6) and (2.7) in Ref.~\cite{Car87}.

For $S^-$ it turns out convenient to choose $r_0^- = 1$, for $S^+$ we choose $r_0^+=0$. With these choices, and applying the transformation (19.22.21) in Ref.~\cite{dlmf} we find
\sbeqs{\eqnlab{spmsol}
\algn{
	S^+(r,s) &= \frac{2}{3}\sqrt{s}\left(1-\frac{{\abs}^2}{4s}\right)(2r)^{3/2}R_D\left(a^2,z_-^2,z_+^2\right) + \sqrt{sr} + \frac{{\abs}}{1-r}\left(\sqrt{Q(r)}+r\right)\,,
	\\
	S^-(r,s) &= \frac23\sqrt{s}\left(1-\frac{{\abs}^2}{4s}\right)\left(1-\sqrt{r}\right)^3R_D(\tilde a^2,\tilde z_-^2,\tilde z_+^2) + \frac{{\abs}}{1-r}	\left(	\sqrt{\frac{Q(r)}{r}}-\sqrt{r}\right)
	\,, 
}
}
with the parameters
\algn{
		a^2 &= 1-r +\frac{{\abs}}{\sqrt{s}}\sqrt{\frac{Q(r)}{r}}+\frac{{\abs}^2 r}{2s}\,,& z^2_\pm &= 1\pm r +\frac{{\abs}}{\sqrt{s}}\sqrt{\frac{Q(r)}{r}}\,,\\
		\tilde a^2 &= \frac{{\abs}^2}{2s}\left(\sqrt{\frac{Q(r)}{r}}+\frac{1+r}2\right)\,,& \tilde z_\pm^2 &= \frac{{\abs}^2}{2s}\left(\sqrt{\frac{Q(r)}{r}}+\sqrt{r}\right)+(1-\sqrt{r})^2(1\pm1)/2\,,}
The next higher orders $T^\pm(r,s)$ obey the equations
\algn{
	T^\pm(r,s) 	&= \int_{r_1^\pm}^r\!\!\ed q \frac{2q(1-3  q)+(1-7 q) (1-q)^2 s/{\abs}^2}{4 (1-q) Q(q)}\mp\int_{r_1^\pm}^r\!\!\ed q \frac{1+q}{2 (1-q) \sqrt{Q(q)}}\,,
}
where $r_1^\pm$ can again be chosen arbitrarily. We decompose the integrands by a partial fraction decomposition. After a series of tedious but straightforward manipulations, using relation (2.15) in Ref.~\cite{Car88} and (19.22.20) in Ref.~\cite{dlmf}, and setting $r_1^-\to1$, we find
\sbeqs{\eqnlab{tpmsol}
\algn{
	T^+(r,s) 	=& \frac{{\abs}}{\sqrt{s}}(r-r_0^+)\left[R_C\left(\frac{\delta^4}{\beta^2}+\frac{{\abs}^2}{s}(r-r_0^+)^2,\frac{\delta^4}{\beta^2}\right) - 2R_C\left(\varphi^2+\frac{{\abs}^2}{s}(r-r_0^+)^2,\varphi^2\right) \right]\nn\\
	&+ \log\left(\frac{1-r}{1-r_0^+}\right) + \frac14\log\left[\frac{Q(r)}{Q(r_0^+)}\right]\,,\\
	T^-(r,s) 	=& \log\left\{\left[\sqrt{Q(r)}+r\right]\frac{\sqrt[4]{ Q(r)}}{2\sqrt{r}}\right\}\,.
}
}	
where $R_C(x,y) \equiv R_F(x,y,y)$ and
\algn{
	\alpha &= \sqrt{r(r_1^+-q_+)(r-q_-)} + \sqrt{r_1^+(r-q_+)(r_1^+-q_-)} \,,\\
	\beta &=  \sqrt{r(r_1^+-q_+)(r_1^+-q_-)}+\sqrt{r_1^+(r-q_+)(r-q_-)}\,,\\
	 \delta^2 &= \alpha^2-(r-r_1^+)^2(1-q_+)\,,\\
	 \varphi^2 &= (1-r)(1-r_1^+)\delta^2\,.
}
The choice $r_0^-,r_1^-\to1$ implies that $S^-\to 0$ and $T^-\to0$ as $r\to1$. From an analysis of the integrand for $S^+$ in \Eqnref{S0pmint} it follows that $S^+\sim 2/(1-r)\to\infty$ as $r\to1$. Combining this with \Eqsref{WKBform} and \eqnref{boundcond} we then find \Eqnref{Aval} in the main text.

Hence, one of the constants of the WKB solution in \Eqnref{WKBform} is immediately fixed by the boundary condition at $r=1$. The result \eqnref{Aval} will be used in the asymptotic matching procedure presented in \secref{bl}.
\section{Solution in boundary layer}\seclab{bl}
Equation~\eqnref{FPrwn} has a boundary layer at $r=0$ which makes the WKB approximation, described in \Secref{ltwkb} and \secref{elliptic}, fail for too small $r$ (or too large ${\abs}$). To see this, we rescale $r$ with ${\loc}$, $r\to {\loc} \tilde r$ and keep only the lowest order terms. This gives
\algn{\eqnlab{FPrbl}
	\partial_{\tilde x}P(\tilde r,\tilde x) \sim \partial_{\tilde r} {\tilde r}(2{\abs} + \partial_{\tilde r})P(\tilde r,\tilde x)\,,
}
with no-flux condition $\lim_{\tilde r\to0}\tilde r(2{\abs} + \partial_{\tilde r})P(\tilde r,\tilde x)=0$. To solve \Eqnref{FPrbl}, we first take a (rescaled) Laplace transform using the coordinate $\tilde s = s/{\loc}$, to obtain
\algn{\eqnlab{FPrlaplace}
	-\delta(1-\tilde r) + \tilde s \hat P(\tilde r,\tilde s) \sim \partial_{\tilde r} \tilde r(2{\abs} + \partial_{\tilde r})\hat P(\tilde r,\tilde s)\,.
}
The initial boundary condition $P(\tilde r,0)=\delta(1-\tilde r)$ corresponds, in the new coordinates, to a delta function outside the boundary layer, so that it can be neglected.

We now write $\hat P(\tilde r,\tilde s) = \exp(-2{\abs}\tilde r)\check P(\tilde r,\tilde s)$. The no-flux condition then translates into the condition $\lim_{r\to0}\tilde r\partial_{\tilde r}\check P(\tilde r,\tilde s)= 0$ for $\check P$ and \Eqnref{FPrlaplace} now reads
\algn{\eqnlab{FPrlaplace2}
\tilde s \check P(\tilde r, \tilde s) = \tilde r\partial_{\tilde r}^2 \check P(\tilde r,\tilde s)  + (1 - 2{\abs}\tilde r)\partial_{\tilde r}\check P(\tilde r,\tilde s)\,,
}
Equation~\eqnref{FPrlaplace2} is the Kummer equation whose independent solutions are the Kummer functions $U(\tilde s/2{\abs},1,2{\abs}\tilde r)$ and $M(\tilde s/2{\abs},1,2{\abs}\tilde r)$~\cite{dlmf}, so that
\algn{\eqnlab{kummersol}
	\check P(\tilde r, \tilde s) \sim A_0 M(\tilde s,1,\tilde r) + B_0 U(\tilde s,1,\tilde r)\,,
}
inside the boundary layer. For $\tilde r\to0$ the Kummer functions behave as~\cite{dlmf}:
\algn{
	U(\tilde s/2{\abs},1,2{\abs}\tilde r) \sim& -\frac1{\Gamma(\tilde s/2{\abs})}\left(\ln (2{\abs}\tilde r) + \psi(\tilde s/(2{\abs})) + 2\Gamma\right)\,,\\
	\quad M(s/(2{\abs}),1,2{\abs} r)\sim&1\,,
}
where $\psi(x)$ denotes the digamma function~\cite{dlmf}. Substituting these expressions into the no-flux condition, we find
\algn{
	\lim_{\tilde r\to 0}\tilde r\partial_{\tilde r}U(\tilde s/(2{\abs}),1,2{\abs}\tilde r) = -\frac1{\Gamma(\tilde s/(2{\abs}))}\,,\quad \lim_{\tilde r\to 0}\tilde r\partial_{\tilde r}M(\tilde s/(2{\abs}),1,2{\abs}\tilde r) = 0\,.
}
This shows that only $M(\tilde s/(2{\abs}),1,2{\abs}\tilde r)$ satisfies the no-flux condition at $\tilde r = 0$, so that $B_0=0$ in \Eqnref{kummersol}. For $\hat P(r,s)$ in the original coordinates we hence obtain
\algn{\eqnlab{blform}
	\hat P(r,s) \sim A_0 \ete^{-2{\abs} r/{\loc}} M(s/(2{\loc\abs}),1,2{\abs} r/{\loc})\,.
}
Finally, this expression needs to be matched to the solution from the WKB approach in \Secref{ltwkb} to obtain the remaining prefactor $A_0$. 
\section{Matching}\seclab{matching}
Here we show how to match the WKB solution in \Eqnref{WKBform}, valid for ${\loc}\ll r\leq1$ to the solution~\eqnref{blform} valid in the boundary layer, for $0\leq r\ll{\loc}$. Matching determines the constant $A_0(s)$ that is needed for computing $\langle \log r\rangle$ in \Eqnref{logrint}.

Recall that we have rescaled $s$ by ${\loc}$ when making the Laplace-transformed WKB ansatz in the beginning of \Secref{ltwkb}. We now undo this rescaling to facilitate matching with the solution inside the boundary layer. We set $(s,r)\to ({\loc} \tilde s,{\loc} \tilde r)$ and expand in ${\loc}$. From \Eqsref{spmsol} and \eqnref{tpmsol} we then obtain for small ${\loc}$
\algn{
	S^+(r,s)/{\loc} + T^+(r,s)\sim& \left(\frac{\tilde s}{{\abs}}+1\right)\log\left(\sqrt{1+\frac{{\abs}^2 \tilde r}{\tilde s}}+\sqrt{\frac{{\abs}^2 \tilde r}{\tilde s}}\right)  +{\abs} \tilde r+\sqrt{\tilde s \tilde r \left(1 + \frac{{\abs}^2 \tilde r}{\tilde s}\right)}\nn\\
	&+ \frac14\log\left({\loc} \tilde r\right)+\frac14\log\left(\frac{{\abs}^2}{{\loc}\tilde s}\right)- \log\left[\left(\sqrt{\frac{\tilde r {\abs}^2}{\tilde s}} + \sqrt{1+\frac{\tilde r {\abs}^2}{\tilde s}}\right)^2-r_0^+\right] \nn\\
	&+ \frac14\log\left(1 + \frac{{\abs}^2 \tilde r}{\tilde s}\right)- \log[(1-r_0^+)^2/2] - \frac12\log\left(r^+_0\right)\,,\\
	S^-(r,s)/{\loc} + T^-(r,s)\sim& - \left(\frac{\tilde s}{{\abs}}-1\right)\log\left[\left(\sqrt{1+\frac{{\abs}^2 \tilde r}{\tilde s}}+\sqrt{\frac{{\abs}^2 \tilde r}{\tilde s}}\right)/2\right] + \frac14\log\left[{\loc}\tilde r{\loc}\tilde s\left(1+\frac{{\abs}^2 \tilde r}{\tilde s}\right)\right]\nn\\
	&-\frac12\log\left({\abs}\right)-\left(\frac{\tilde s}{2{\abs}}-\frac12\right)\log\left(\frac{{\loc}\tilde s}{{\abs}^2}\right) + \frac{\tilde s}{2{\abs}} +{\abs}\tilde r - \sqrt{\tilde r\tilde s\left(1+\frac{{\abs}^2 \tilde r}{\tilde s}\right)}
}
From these expressions, we observe that for large $\tilde r$, $\ete^{-S^+/{\loc} - T^+}$ is exponentially suppressed compared to $\ete^{-S^-/{\loc} - T^-}$, so that only $\ete^{-S^-/{\loc} - T^-}$ can match $\hat P(r,s)$ in \Eqnref{blform}.

We use the uniform asymptotic expansion in given Ref.~\cite{Tem22} to determine the asymptotic form of $\hat P(r,s)$ in the boundary layer [\Eqnref{blform}] as ${\loc}\to0$, We find
\algn{
	\hat P(r,s) \sim \frac{A_0}{\Gamma\left(\frac{s}{2{\loc\abs}}\right)}\left(\frac{s}{2{\loc\abs}}\right)^{\frac{s}{2{\loc\abs}}-\frac12}\sqrt{\frac{{\loc}}2}\ete^{-\frac{s}{2{\loc\abs}}}\frac{\left(\sqrt{\frac{{\abs}^2 r}{s}}+\sqrt{1+\frac{{\abs} ^2 r}{s}}\right)^{\frac{s}{{\loc\abs}}-1}}{\sqrt[4]{rs \left(1+\frac{{\abs} ^2 r}{s}\right)}}\ete^{-\frac1{{\loc}}\left[{\abs} r+\sqrt{rs \left(1+\frac{{\abs} ^2 r}{s}\right)}\right]}\,.
}
This must match with the WKB solution in the boundary layer
\algn{
	\hat P(r,s)	\sim \frac{A_-}{\sqrt{2\pi{\loc}}}\left(\frac{s}{4{\abs}^2}\right)^{\frac{s}{2{\loc\abs}}-\frac12}\sqrt{{\abs}}\ete^{-\frac{s}{2{\loc\abs}} }\frac{\left(\sqrt{1+\frac{{\abs}^2 r}{s}}+\sqrt{\frac{{\abs}^2 r}{s}}\right)^{\frac{s}{{\loc\abs}}-1}}{\sqrt[4]{r s\left(1+\frac{{\abs}^2 \tilde r}{\tilde s}\right)}}		\ete^{-\frac1{{\loc}}\left[{\abs} r + \sqrt{rs\left(1+\frac{{\abs}^2r}{s}\right)}\right]}\,.
}
With $A_-/\sqrt{2\pi{\loc}}=1/(2{\abs})$ [see \Eqnref{Aval}]. We find that matching is achieved when
\algn{
	\frac{A_0(s)}{\Gamma\left(\frac{s}{2{\loc\abs}}\right)}\left(\frac{s}{2{\loc\abs}}\right)^{\frac{s}{2{\loc\abs}}-\frac12}\sqrt{\frac{{\loc}}2} = \frac1{2\sqrt{{\abs}}}\left(\frac{s}{4{\abs}^2}\right)^{\frac{s}{2{\loc\abs}}-\frac12}\,,
}
which gives \Eqnref{A0expr} for $A_0(s)$ in the main text.
\section*{References}
\providecommand{\newblock}{}

\end{document}